\documentclass[chap,11pt]{usthesis}
\usepackage{epsfig,amssymb,multirow,amsmath,enumerate}
\usepackage{hyperref}

\begin{document}

\newcommand\beq{\begin{equation}}      
\newcommand\beqnn{\begin{eqnarray*}}   
\newcommand\beqa{\begin{eqnarray}}     
\newcommand\beqann{\begin{eqnarray*}}  

\newcommand\eeq{\end{equation}}        
\newcommand\eeqnn{\end{eqnarray*}}     
\newcommand\eeqa{\end{eqnarray}}       
\newcommand\eeqann{\end{eqnarray*}}    

\newcommand{\diffpa}[1]{\frac{\partial}{\partial {#1}}}                                
\newcommand{\diff}[2]{\frac{\textnormal{d} {#1}}{\textnormal{d} {#2}}}                      
\newcommand{\diffpabrack}[2]{\frac{\partial}{\partial {#2}} \left( {#1} \right)}            
\newcommand{\diffbrack}[2]{\frac{\textnormal{d}}{\textnormal{d} {#2}} \left( {#1} \right)}  

\newcommand{\ket}[1]{\left| #1 \right\rangle}                                               
\newcommand{\bra}[1]{\left\langle #1\right|}                                               
\newcommand{\ketr}[1]{\left| #1 \right)}                                                    
\newcommand{\brar}[1]{\left( #1\right|}                                                    
\newcommand{\overlap}[2]{\left\langle {#1} | {#2} \right\rangle}                            
\newcommand{\matrel}[3]{\left\langle {#1} \right| {#2} \left| {#3}\right\rangle}            
\newcommand{\overlapr}[2]{\left( {#1} | {#2} \right)}                                       
\newcommand{\matrelr}[3]{\left( {#1} \right| {#2} \left| {#3}\right)}                       

\def\bd {b^{\dagger}}
\def\bl {B_{L}}
\def\bld {B_{L}^{\ddagger}}
\def\br {B_{R}}
\def\brd {B_{R}^{\ddagger}}
\def\zb {\bar{z}}
\def\vb {\bar{v}}
\def\pb {\bar{p}}
\def\qb {\bar{q}}
\def\hq {\mathcal{H}_q}
\def\hc {\mathcal{H}_c}

\newcommand{\tr}[1]{\textnormal{tr}_{c} \left( #1 \right) }                                 
\newcommand{\func}[2]{\textnormal{#1}\left( #2 \right)}                                     
\newcommand{\mb}[1]{\mathbf{#1}}                                                            
\newcommand{\mtn}[1]{\textnormal{#1}}                                                       
\newcommand{\modsq}[1]{\left| #1 \right|^{2}}                                               
\def\nl {\nonumber \\}
\def\etal{\emph{et al. }}

\let\oldsqrt\sqrt
\def\sqrt{\mathpalette\DHLhksqrt}
\def\DHLhksqrt#1#2{%
\setbox0=\hbox{$#1\oldsqrt{#2\,}$}\dimen0=\ht0
\advance\dimen0-0.2\ht0
\setbox2=\hbox{\vrule height\ht0 depth -\dimen0}%
{\box0\lower0.4pt\box2}}

\thesistitle{ADDITIONAL DEGREES OF FREEDOM ASSOCIATED WITH POSITION MEASUREMENTS IN NON-COMMUTATIVE QUANTUM MECHANICS}
\author{C. M. ROHWER}
\degree{Master~of~Science}
\supervisor{Professor F.G. Scholtz}
\submitdate{September 2010}

\titlepage
\declaration

\specialhead{Abstract}

Due to the minimal length scale induced by non-commuting co-ordinates, it is not clear \emph{a priori} what is meant by a position measurement on a non-commutative space. It was shown recently in a paper by Scholtz \etal that it is indeed possible to recover the notion of quantum mechanical position measurements consistently on the non-commutative plane. To do this, it is necessary to introduce weak (non-projective) measurements, formulated in terms of Positive Operator-Valued Measures (POVMs). In this thesis we shall demonstrate, however, that a measurement of position alone in non-commutative space cannot yield complete information about the quantum state of a particle. Indeed, the aforementioned formalism entails a description that is non-local in that it requires knowledge of all orders of positional derivatives through the star product that is used ubiquitously to map operator multiplication onto function multiplication in non-commutative systems. It will be shown that there exist several equivalent local descriptions, which are arrived at via the introduction of additional degrees of freedom. Consequently non-commutative quantum mechanical position measurements necessarily confront us with some additional structure which is necessary (in addition to position) to specify quantum states completely. The remainder of the thesis, based in part on a recent publication (\textbf{\emph{``Noncommutative quantum mechanics -- a perspective on structure and spatial extent''}, C.M. Rohwer, K.G. Zloshchastiev, L. Gouba and F.G. Scholtz, J. Phys. A: Math. Theor. \textbf{43} (2010) 345302}) will involve investigations into the physical interpretation of these additional degrees of freedom. For one particular local formulation, the corresponding classical theory will be used to demonstrate that the concept of extended, structured objects emerges quite naturally and unavoidably there. This description will be shown to be equivalent to one describing a two-charge harmonically interacting composite in a strong magnetic field found by Susskind. It will be argued through various applications that these notions also extend naturally to the quantum level, and constraints will be shown to arise there. A further local formulation will be introduced, where the natural interpretation is that of objects located at a point with a certain angular momentum about that point. This again enforces the idea of particles that are not point-like. Both local descriptions are convenient, in that they make explicit the additional structure which is encoded more subtly in the non-local description. Lastly we shall argue that the additional degrees of freedom introduced by local descriptions may also be thought of as gauge degrees of freedom in a gauge-invariant formulation of the theory.

\specialhead{Opsomming}

As gevolg van die minimum lengteskaal wat deur nie-kommuterende ko\"ordinate ge\"induseer word is dit nie \emph{a priori} duidelik wat met 'n posisiemeting op 'n nie-kommutatiewe ruimte bedoel word nie. Dit is onlangs in 'n artikel deur Scholtz \etal getoon dat dit wel op 'n nie-kommutatiewe vlak  moontlik is om die begrip van kwantummeganiese posisiemetings te herwin. Vir hierdie doel benodig ons die konsep van swak (nie-projektiewe) metings wat in terme van 'n positief operator-waardige maat geformuleer word. In hierdie tesis sal ons egter toon dat 'n meting van slegs die posisie nie volledige inligting oor die kwantumtoestand van 'n deeltjie in 'n nie-kommutatiewe ruimte lewer nie. Ons formalisme behels 'n nie-lokale beskrywing waarbinne kennis oor alle ordes van posisieafgeleides in die sogenaamde sterproduk bevat word. Die sterproduk is 'n welbekende konstruksie waardeur operatorvermenigvuldiging op funksievermenigvuldiging afgebeeld kan word. Ons sal toon dat verskeie ekwivalente lokale beskrywings bestaan wat volg uit die invoer van bykomende vryheidsgrade. Dit beteken dat nie-kommutatiewe posisiemetings op 'n natuurlike wyse die nodigheid van bykomende strukture uitwys wat noodsaaklik is om die kwantumtoestand van 'n sisteem volledig te beskryf. Die res van die tesis, wat gedeeltelik op 'n onlangse publikasie (\textbf{\emph{``Noncommutative quantum mechanics -- a perspective on structure and spatial extent''}, C.M. Rohwer, K.G. Zloshchastiev, L. Gouba and F.G. Scholtz, J. Phys. A: Math. Theor. \textbf{43} (2010) 345302}) gebaseer is, behels 'n ondersoek na die fisiese interpretasie van hierdie bykomende strukture. Ons sal toon dat vir 'n spesifieke lokale formulering die beeld van objekte met struktuur op 'n natuurlike wyse in die ooreenstemmende klassieke teorie na vore kom. Hierdie beskrywing is inderdaad ekwivalent aan die van Susskind wat twee gelaaide deeltjies, gekoppel deur 'n harmoniese interaksie, in 'n sterk magneetveld behels. Met behulp van verskeie toepassings sal ons toon dat hierdie interpretasie op 'n natuurlike wyse na die kwantummeganiese konteks vertaal waar sekere dwangvoorwaardes na vore kom. 'n Tweede lokale beskrywing in terme van objekte wat by 'n sekere punt met 'n vaste hoekmomentum gelokaliseer is sal ook ondersoek word. Binne hierdie konteks sal ons weer deur die begrip van addisionele struktuur gekonfronteer word. Beide lokale beskrywings is gerieflik omdat hulle hierdie bykomende strukture eksplisiet maak, terwyl dit in die nie-lokale beskrywing deur die sterproduk versteek word. Laastens sal ons toon dat die bykomende vryheidsgrade in lokale beskrywings ook as ykvryheidsgrade van 'n ykinvariante formulering van die teorie beskou kan word.


\specialhead{Acknowledgements}

I would like to express my sincerest thanks to my supervisor, Professor F.G. Scholtz. Due to the interpretational slant of this thesis, there were frequently periods where the path forward was unclear. For his accommodating support in identifying sensible questions and for his open door when the corresponding answers were evasive, I am most grateful.

The Wilhelm Frank Bursary Trust, administrated by the Department of Bursaries and Loans at Stellenbosch University, provided financial support for my studies, not only during my M.Sc. but also during my B.Sc. and B.Sc. Hons. degrees. This aid was instrumental in allowing me to focus on academic priorities and I am thankful for the privilege of having received it.

The many conversations with my fellow students A. Hafver and H.J.R. van Zyl were of immeasurable value and helped to shed light on several matters. I appreciate not only the academic aspects of these exchanges, but also the friendship and sound-board for voicing frustrations.

A great word of thanks is due to Mr. J.N. Kriel for helping me with technical and calculational issues and for engaging in countless interpretational discussions. His many hours of patient and involved assistance are valued sincerely.

For their hospitality at the S.N. Bose National Centre for Basic Sciences, Kolkata, and at the Centre for High Energy Physics of the Indian Institute of Science, Bangalore, I am most grateful to Professors B. Chakraborty and S. Vaidya, respectively. The visit to India during 2010 provided a valued forum for exchanging ideas and identifying interesting problems for future work.

Lastly, and perhaps most importantly, I would like to thank my parents for their unconditional support and love. Their patience and understanding mean a great deal to me.

\tableofcontents
\listoffigures
\specialhead{Background and Motivations}

As strange as the idea of introducing non-commutative spatial co-ordinates into quantum mechanical theories may seem, it is certainly not as novel as one may expect. In fact, suggestions that space-time co-ordinates may be non-commutative appeared already in the early days of quantum mechanics. For instance, in his article \cite{snyder} of 1947, Snyder pointed out that it is problematic to describe matter and local interactions relativistically in continuous 4-dimensional space-time due to the appearance of divergences in field theories in this context. It is shown there that there exists Lorentz-invariant space-time in which there is a natural unit of length, the introduction of which partially remedies aforementioned divergences. It is also demonstrated that the notion of a smallest unit of length can only be implemented upon having dropped assumptions of commutative space-time: commuting co-ordinates would have continuous spectra which would contradict the idea of spatial quantisation. More recently, the notion of non-commutative space-time was investigated also from the perspective of gravitational instabilities. In \cite{dopplicher}, Dopplicher \etal argued that attempts at spatial localisation with precision smaller than the Planck length,
\beq
\ell_p= \sqrt {\frac{G\hbar}{c^3}} \simeq 1.6\times10^{-33}\;\textnormal{cm},
\label{lp}
\eeq
result in the collapse of gravitational theories in that they would require energy concentrations large enough to induce black hole formation. A natural solution to this problem would be to impose a minimum bound on localisability. On an intuitive level, one may understand this to be a consequence of the fact that a minimal length scale implies a regularisation of high momenta (through the Fourier transformation), which in turn restricts the attainable energy concentrations. Since non-commuting operators induce uncertainty relations, a natural way to impose such a bound on localisability is to introduce co-ordinates that do not commute. By finding a Hilbert space representation of a non-commuting algebra, these authors then introduced the concept of optimal localisation and put forward first steps toward field theories in this context. 

At this point, already, one may ask whether the notion of a point particle makes any sense in a space with finite, non-zero minimal bounds on spatial localisability. Though the answer to this question is far from obvious, it is clear that a local description of a point particle, i.e., one where we allocate a specific position to a point particle which has no physical extent, is nonsensical if we cannot specify co-ordinates to arbitrary accuracy. One fundamental motivation behind the study of frameworks such as string theory, is the need for a consistent field theoretical framework for extended objects where the notion of point-like local interactions may be replaced by non-local interactions. In this particular context, it was shown by Susskind \etal that a free particle moving in the non-commutative plane can be thought of as two oppositely charged particles interacting through a harmonic potential and moving in a strong magnetic field \cite{susskind} --- an idea which makes the notion of physical extent and structure quite explicit. This article also alludes to the important role that non-commutative geometries play in the framework of string theory. Seiberg explains in \cite{seiberg} that the extended nature of strings leads to ambiguities in defining geometry and topology of space-time, and that field theories on non-commutative space in fact correspond to low-energy limits of string theories. Indeed, the study of field theories on non-commutative geometries -- another setting in which non-local interactions occur quite naturally --  has grown into a sizeable research field of its own; for an extensive review, see, for instance, \cite{douglas}. Furthermore, the framework of non-commutative geometry provides a useful mathematical setting for the study of matrix models in string theory, as set out in \cite{connes}. In the context of the Landau problem, non-commutativity of guiding center co-ordinates in the lowest Landau level is well-known (the non-commutative parameter here scales inversely to the magnitude of the magnetic field); a detailed discussion can be found in \cite{jain}. Thermodynamic quantities such as the entropy of a non-commutative fermion gas have also been shown to exhibit non-extensive features due to the excluded volume effects induced at high densities by non-commutativity \cite{thermodynamics}. Non-commutative geometry appears in various other physical applications -- a comprehensive summary may be found in \cite{ncgeometry}.

\hbox{}
Returning to the issues discussed earlier in this chapter, we see from many arguments there that the standard views of space-time merit further scrutiny and possibly even drastic revision. Evidently one candidate for addressing many of the problems encountered in this context is the introduction of non-commutative spatial co-ordinates. We have also seen that the issues of locality of measurements and the notion of extendedness go hand-in-hand with such modified space-time frameworks. Indeed, a consistent probability framework to describe position measurements in non-commutative quantum mechanics --- a matter which is not trivial since non-commuting co-ordinates do not allow for simultaneous eigenstates --- was formulated in \cite{jopa}. This thesis departs with a detailed investigation of the non-locality\footnote{ With non-locality we mean that this description requires knowledge not only of the position wave function, but also of all orders of spatial derivatives thereof. This non-locality is encoded in the so-called star product, and will be elaborated upon in the chapters to follow.} of this description. We shall then proceed to introduce a manifestly local description for non-commutative quantum mechanical position measurements on a generic level, and subsequently focus on two specific choices of basis and their interpretations. As stated above, local position measurements of point particles do not make sense in non-commutative space. Consequently it is not surprising that aforementioned local descriptions require the introduction of additional degrees of freedom and that constraints arise in some of these formulations. At this point it would be quite natural to ask whether non-commutative quantum mechanics might allow for an interpretation in terms of objects with additional structure and / or extent. Indeed, it has been shown (see Section 2 of \cite{chris}) that the conserved energy and total angular momentum derived from the non-commutative path integral action in \cite{sunandan} contain explicit correction terms to those for a point particle. Thus we see that, already on a classical level, there are hints at structured objects in a non-commutative theory. To provide further motivation for this standpoint, we shall show that the physical picture of Susskind \cite{susskind} that was mentioned above appears explicitly in the non-commutative classical theory corresponding to one of our local descriptions. In this context we shall also demonstrate that the aforementioned correction terms to the conserved energy may also be formulated in terms of the additional degrees of freedom of this local description. Through application of this formulation to eigenstates of angular momentum, the free particle and the quantum harmonic oscillator we shall demonstrate that Susskind's view is natural also in the context of non-commutative quantum mechanics. A further local formulation will be shown to allow a natural interpretation in terms of objects with an angular momentum about a point of localisation. Naturally such a point of view is incompatible with that of a point-particle whose internal degrees of freedom have not been specified. We shall conclude that the notion of additional structure is undeniably present in any such local description, and, most importantly, that complete information about non-commutative quantum mechanical states cannot be provided by a measurement of position only.

\chapter{A REVIEW OF THE STANDARD QUANTUM MECHANICAL FRAMEWORK}

In the following two chapters we shall discuss in detail the formalism that will be used for the remainder of this thesis. In order to illustrate the consequences of introducing non-commutative co-ordinates, we first review standard quantum mechanics, focusing on the significance of algebraic commutation relations and the statistical interpretations of measurement processes. Thereafter the non-commutative formalism as set out in \cite{jopa} will be introduced and described in detail, with particular attention payed to the identification of measurable quantities and a suitable framework for position measurements. Note that all analyses will be restricted to two dimensions, i.e., our formalism applies to a non-commutative plane.\footnote{ In Appendix \ref{thirdcoord} we discuss briefly the inclusion of a third co-ordinate and the associated problems pertaining to transformation properties and rotational invariance in higher dimensions.}

\section{A unitary representation of the Heisenberg algebra}
The cornerstone of standard quantum mechanics is the set of canonical commutation relations. The relevant underlying structure is the abstract Heisenberg algebra, which reads
\beqa
\label{heisc}
\left[{x},{y}\right] &=& 0,\nonumber\\
\left[{x},{p}_x\right] = \left[{y},{p}_y\right] &=& i\hbar,\\
\left[{p}_x,{p}_y\right]= \left[x,{p}_y\right] = \left[y,{p}_x\right]&=& 0\nonumber.
\eeqa
The generators of the algebra are linked to observable quantities through the construction of a unitary representation in terms of Hermitian operators that act on the quantum mechanical Hilbert space. The states of the system are represented by vectors in this quantum Hilbert space, which shall henceforth be denoted by $\hq$. These operators obviously obey the same commutation relations as those above,
\beqa
\label{heisc2}
\left[\hat{x},\hat{y}\right] &=& 0,\nonumber\\
\left[\hat{x},\hat{p}_x\right] = \left[\hat{y},\hat{p}_y\right] &=& i\hbar,\\
\left[\hat{p}_x,\hat{p}_y\right]= \left[\hat{x},\hat{p}_y\right] = \left[\hat{y},\hat{p}_x\right]&=& 0\nonumber.
\eeqa
Two representations are common in the setting of standard quantum mechanics --- the Schr\"odinger representation and Heisenberg's matrix representation.\footnote{ From the \emph{Stone-von Neumann} theorem we know that all unitary representations of the algebra (\ref{heisc}) are equivalent; see, for instance, \cite{stonevonneumann}.} For the former, for instance, we have that the position and momentum operators act on the Hilbert space of square-integrable wave functions as follows:
\beqa
\hat{x}\psi(x,y) &=& x\psi(x,y),\nl
\hat{p}_x\psi(x,y) &=& -i\hbar \frac{\partial}{\partial x}\psi(x,y),
\label{schrodrep}
\eeqa
and similarly for $y$. \\ \\
The commutation relations (\ref{heisc2}) induce an uncertainty in the position and momentum observables,
\beq
\Delta \hat{x}\Delta \hat{p}_x \geq \frac{\hbar}{2}, \quad \Delta \hat{y}\Delta \hat{p}_y \geq \frac{\hbar}{2},
\label{pxuncert}
\eeq
where we define $\Delta \hat{A} \equiv \sqrt{\langle \hat{A}^2 \rangle - \langle \hat{A} \rangle^2 }$ for any observable $\hat{A}$.\footnote{ The proof hereof is simple, and relies on the Schwartz inequality; see, for instance, \cite{sakurai}.} On a physical level, (\ref{pxuncert}) simply implies that, for a given direction, momentum and position cannot be measured simultaneously to arbitrary accuracy. In contrast to this, however, the two co-ordinates may be measured simultaneously since $x$ and $y$ commute in (\ref{heisc}), as do the operators (\ref{heisc2}) representing them on the Hilbert space. It is this particular feature that will later be altered in a non-commutative setting. \\

Having reviewed the matter of representations of the abstract Heisenberg algebra, let us revisit the statistical interpretation associated with measurements in the standard quantum mechanical formalism.

\section{The postulates of standard quantum mechanics}

In standard quantum mechanics, measurements are considered to be projective. To illustrate precisely what is meant by this statement, we now recap the fundamental postulates of this probabilistic framework. We shall follow the discussion of \cite{bergou}, where the quantum mechanical formulation of von Neumann (see, for instance, \cite{neumann}) is summarised. \\

\textbf{Postulates of Standard Quantum Mechanics:}
\begin{enumerate}[I]
  \item{To every quantum mechanical observable we assign a corresponding Hermitian operator, $A=A^\dagger$. Due to the Hermiticity of $A$, we can construct a complete orthonormal basis (which, for simplicity, we assume here to be discrete) for $\hq$ from the eigenvectors of $A$:
      \beq
      A\ket{\phi_n}=\lambda_n\ket{\phi_n},\;\overlap{\phi_n}{\phi_m}=\delta_{n,m}\quad \Rightarrow \quad\hq = \textnormal{span}_n\{\ket{\phi_n}\}.\footnote{ At this point we do not stipulate the dimensionality of $\hq$.}
      \label{sp1}
      \eeq
      Naturally $A$ has a spectral representation in terms of these eigenvectors:
      \beqa
      A &=& \sum_n  \lambda_n \ket{\phi_n}\bra{\phi_n} \nl
      &\equiv& \sum_n  \lambda_n P_n.
      \label{sp2}
      \eeqa
      The Hermiticity of $A$ also guarantees a real spectrum, $\lambda_n \in \Re \;\;\forall\; n$.}

  \item {We call the operators $P_n \equiv \ket{\phi_n}\bra{\phi_n}$ projectors. They sum to the identity on the quantum Hilbert space,
      \beq
      \sum_n  P_n = \mathbf{1}_q.
      \label{sp3}
      \eeq

      Since the eigenvectors of $A$ are orthogonal (see (\ref{sp1})), we have that
      \beq
      P_nP_m = \ket{\phi_n}\overlap{\phi_n}{\phi_m}\bra{\phi_m}=\delta_{n,m}P_n.
      \label{sp6}
      \eeq
      Consequently any projector squares to itself, i.e., $P_n^2 = P_n$. This implies that its eigenvalues must be $0$ or $1$.}
  \item A measurement of the observable $A$ must necessarily yield one of the eigenvalues of $A$, say $\lambda_\alpha \in \{\lambda_n\;|\;n=0:\infty\}$. If the system is originally in a normalised pure state $\ket{\psi}$, then the probability\footnote{ The inherent randomness in the measurement process becomes manifest in this postulate: we are not guaranteed any particular outcome. The only prediction we can make is the set of possible outcomes, and to each element thereof we may assign a probability. It is in this context that the notion of ensemble measurements is a natural interpretation.} of measuring $\lambda_\alpha$ is given by
      \beq
      p_\alpha =\modsq{\overlap{\phi_\alpha}{\psi}}=\matrel{\psi}{P_\alpha}{\psi} =\matrel{\psi}{P_\alpha^2}{\psi}=\modsq{P_\alpha\ket{\psi}}.
      \label{sp4}
      \eeq
      These probabilities are non-negative and sum to unity,

      \beq
      p_\alpha \geq 0, \; \sum_{\alpha} p_\alpha = 1 \quad \Rightarrow \quad  0 \leq p_\alpha \leq 1,
      \label{sp7}
      \eeq
      as is required for any probability. (These statements are easily verified using equations (\ref{sp2}), (\ref{sp3}) and (\ref{sp4})).

      The normalised state of the system after measurement is
      \beq
      \ket{\phi} \equiv \frac{P_\alpha\ket{\psi}}{\sqrt{\matrel{\psi}{P_\alpha}{\psi}}}.
      \label{sp8}
      \eeq
      If another measurement is performed immediately on the system, it is clear from (\ref{sp2}) and (\ref{sp6}) that the outcome will again be $\lambda_\alpha$ with a probability of 1. It is in this sense that we consider measurements to be \textbf{projective}, since repeated measurements of a particular observable will yield the same result, i.e., the system is projected into a particular eigenstate of the observable in the measurement process.

      For the case where the system is initially in a mixed state described by the density operator $\rho$, the probability of measuring outcome $\lambda_\alpha$ is
      \beq
      p_\alpha = \textnormal{tr}_q (P_\alpha \rho P_\alpha)=\textnormal{tr}_q (P_\alpha^2 \rho )=\textnormal{tr}_q (P_\alpha \rho),
      \eeq
      and the corresponding post-measurement state is described by the density operator
      \beq
      \rho_\alpha =\frac{P_\alpha \rho P_\alpha}{\textnormal{tr}_q (P_\alpha \rho P_\alpha)}=\frac{P_\alpha \rho P_\alpha}{\textnormal{tr}_q (P_\alpha \rho)}.
      \eeq

      (Here $\textnormal{tr}_q$ denotes the trace over the quantum Hilbert space, $\hq$).

  \item The expectation value of $A$, in the sense of repeated measurements on an ensemble of identically prepared systems initially in state $\ket{\psi}$, is given by the probability-weighted sum of all possible outcomes,
      \beq
      \langle A \rangle = \sum_n  p_n \lambda_n.
      \eeq
      The extension to a system initially in the mixed state $\rho$ is simply
      \beq
      \langle A \rangle = \sum_n  \lambda_n\textnormal{tr}_q\{P_n \rho\} = \textnormal{tr}_q\{A\rho\}.
      \eeq

\end{enumerate}

\hbox{}
\begin{noindent}
Although the above postulates outline the usual approach to / interpretation of the statistical quantum mechanical framework, it is possible to relax some of these points. Indeed, the stipulation of projectivity in measurements is a very restrictive one, and we shall demonstrate in the following section that it is possible to build a consistent probabilistic framework where this requirement is relaxed.
\end{noindent}

\section{Weak measurement: the language of Positive Operator Valued Measures}
\label{POVMsection}
One of the most important underpinnings of quantum mechanics is the conservation of probability. This is guaranteed by insisting on Hermiticity of observables, which ensures unitarity in dynamic evolution of the system. Naturally such a description must be applied to closed quantum systems which are devoid of interactions with an environment that may violate conservation. Indeed, open quantum systems are typically described in terms of non-Hermitian operators that represent coupling to the environment.\footnote{ For an example of such a description, see \cite{prosen}.} Generally such descriptions involve an alteration of the postulates set out above. As will be seen in later sections, it is necessary also in the framework of non-commutative quantum mechanical position measurements to modify the postulates of measurement slightly. For this reason we shall consider here a well-established extension to the statistical formalism above, that is of use not only in our framework but also in fields like quantum computing \cite{preskill} and open quantum systems \cite{davies}.

Returning to the matter at hand, we note that, to build a consistent probability framework, it is necessary to have a set of non-negative normalised probabilities as in (\ref{sp7}). Looking at equation (\ref{sp4}), we note that this is possible even if the operators $P_n$ are not positive: it suffices to have positivity for $P_n^2$. We will show that this can be done even if one abandons the requirement of orthogonality (\ref{sp6}) for the operators $P_n$ which generate the post-measurement state (\ref{sp8}).

Suppose now that the normalised post-measurement state after a specific experiment,
\beq
\ket{\phi} = \frac{D_\alpha \ket{\psi}}{\sqrt{\matrel{\psi}{D_{\alpha}^\dagger D_\alpha}{\psi}}},
\label{pms} 
\eeq
is determined by a set of \emph{non-orthogonal} operators \{$D_n$\}, $D_n D_m \neq \delta_{n,m}$. We call these operators ``detection operators'', and they are a generalisation of the orthogonal projectors $P_n$ from (\ref{sp2}). As an extension of the operators $P_n^2$, we further introduce a set of positive operators $\pi_n$ that sum to the identity on $\hq$,
\beq
\pi_n \geq 0, \quad \sum_n  \pi_n=\mathbf{1}_q.
\eeq
With this we have a so-called Positive Operator Valued Measure (POVM), where each $\pi_\alpha$ is an element of the POVM. We note that one way to guarantee positivity of the POVM elements is through the identification
\beq
\pi_n= D_n^\dagger D_n.
\label{Pin}
\eeq
The obvious choice of detection operator would be $D_n = \pi_n^{1/2}$ (the square root of $\pi_n$ exists since the operator is positive). However, the most general choice of detection operator satisfying (\ref{Pin}) is
\beq
D_n = U_n \pi_n^{1/2},
\eeq
where $U_n$ is an arbitrary unitary transformation whose relevance will become clear shortly. Comparing the post-measurement states (\ref{sp8}) and (\ref{pms}), we note that it would be natural to associate the $\alpha$ in $D_\alpha$ with a particular outcome of an observable (which need not be Hermitian). Let us proceed by introducing a modified set of postulates (based on the above POVMs) that allows the construction of a consistent, non-projective quantum mechanical probability interpretation for measurements of such quantities.\\

\textbf{Modified Postulates of Quantum Mechanics: Non-Projective Measurements}
\label{newpostulates}
\begin{enumerate}[I]
\item We no longer require that the operators representing observables  on $\hq$ need be Hermitian. (This need not imply that Hermitian observables no longer exist, we simply do not demand Hermiticity of all observables).
\item Our point of departure is a decomposition of the identity on $\hq$ in terms of positive operators (i.e. a POVM):
    \beq
    \pi_n \geq 0 \;\;\forall\; n, \quad \sum_n  \pi_n=\mathbf{1}_q.
    \eeq
    The elements of the POVM may be decomposed further in terms of so-called detection operators,
    \beq
    \pi_n= D_n^\dagger D_n,
    \eeq
    where $D_n^\dagger \neq D_n$ and $D_n D_m \neq \delta_{n,m}$ in general, but where
    \beq
    \sum_n  D_n^\dagger D_n=\mathbf{1}_q.
    \eeq

\item A detection must necessarily yield an outcome corresponding to one of the elements of the POVM, say $\pi_\alpha$. If the system is originally in a normalised pure state $\ket{\psi}$, then the probability of this particular outcome is
    \beq
    p_\alpha = \modsq{D_\alpha\ket{\psi}}=\matrel{\psi}{D_\alpha^\dagger D_\alpha}{\psi}=\matrel{\psi}{\pi_\alpha}{\psi}.
    \eeq
    The previous postulate ensures that $p_\alpha \geq 0$, $\sum_{\alpha} p_\alpha = 1$, and consequently that $0 \leq p_\alpha \leq 1$.
    For a system initially in a mixed state with density operator $\rho$, the probability for this particular outcome is
    \beq
    p_\alpha=\textnormal{tr}_q (D_\alpha \rho D_\alpha^\dagger)=\textnormal{tr}_q (D_\alpha^\dagger D_\alpha  \rho )=\textnormal{tr}_q (\pi_\alpha\rho ).
    \eeq
    Note that we do not prescribe the form of the operator corresponding to the observable quantity. Rather we consider the set of possible outcomes of a measurement, each being a label of a particular POVM. It need not be the case that these outcomes are necessarily eigenstates of a particular operator.
\item The state of the system after measurement is
    \beq
    \ket{\phi} = \frac{D_\alpha \ket{\psi}}{\sqrt{\matrel{\psi}{D_{\alpha}^\dagger D_\alpha}{\psi}}}.
    \label{pms2}
    \eeq
    Recalling that the most general form of the detection operators is
    \beq
    D_n = U_n \pi_n^{1/2},
    \eeq
    we see that the state (\ref{pms2}) can only be specified up to a unitary transformation, which induces a degree of arbitrariness after measurement. Consequently we cannot make any exact statements about the post-measurement state other than its norm. Furthermore, due to the non-orthogonality of the detection operators, a repeated measurement need not yield the same result (in contrast to (\ref{sp8})). It is in this sense that this framework describes \textbf{non-projective} measurements.

    For a mixed state $\rho$, the post-measurement state of the system is described by
    \beq
    \rho_\alpha =\frac{D_\alpha \rho D_\alpha^\dagger}{\textnormal{tr}_q (D_\alpha \rho D_\alpha^\dagger)}=\frac{D_\alpha \rho D_\alpha^\dagger}{\textnormal{tr}_q (D_\alpha^\dagger D_\alpha \rho )}.
    \eeq
\item For any observable $O$, the expectation value is defined as
    \beq
    \langle O\rangle \equiv\textnormal{tr}_q(O\rho).
    \eeq
\end{enumerate}

\hbox{}
\begin{noindent}
This concludes our review of the standard quantum mechanical framework and the associated probabilistic formalism(s) for describing measurements. We now introduce the non-commutative framework by modifying the Heisenberg algebra (\ref{heisc}), finding a suitable unitary representation on $\hq$, and discussing the implications of this formalism on position measurements.
\end{noindent}

\chapter{THE FORMALISM OF NON-COMMUTATIVE QUANTUM MECHANICS}
\label{ncformalism}
\section{A unitary representation of the non-commutative Heisenberg algebra}
The framework of non-commutative quantum mechanics that we will use was put forward in \cite{jopa}, where a consistent probability interpretation for this formalism was outlined. Said article essentially comprises a consolidation and subsequent extension of the basic machinery used in \cite{torus} and \cite{ncwell}. Since this construction is vital to our analyses, we shall review these discussions thoroughly.

The foundation of our construction is the introduction of a non-commutative configuration space.\footnote{ We will investigate the case where the commutation relations of the momenta are unchanged, and only positional commutation relations are altered.} Co-ordinates on this space satisfy the commutation relation
\begin{equation}
[\hat{x}, \hat{y}] = i\theta,
\label{xycommutator}
\end{equation}
where  $\theta$ is a real parameter (in units of length squared) that is assumed to be positive without loss of generality. By implication, the first line of the abstract Heisenberg algebra (\ref{heisc}) is modified; its non-commutative analogue reads
\beqa
\label{heisnc}
\left[{x},{y}\right] &=& i\theta,\nonumber\\
\left[{x},{p}_x\right] = \left[{y},{p}_y\right] &=& i\hbar,\\
\left[{p}_x,{p}_y\right]= \left[x,{p}_y\right] = \left[y,{p}_x\right]&=& 0\nonumber.
\eeqa
The task at hand is to find the quantum Hilbert space, $\hq$, and a unitary representation of the non-commutative algebra on this space. Returning to (\ref{xycommutator}), we note that non-commutative co-ordinates cannot be scalars since these would commute. For this reason we denote the co-ordinates by hatted operators in (\ref{xycommutator}). In order to find a basis for classical configuration space, it is convenient to define the following creation and annihilation operators:

\begin{eqnarray}\nonumber
b &=& \frac{1}{\sqrt{2\theta}} (\hat{x}+i\hat{y}),\\
b^\dagger &=&\frac{1}{\sqrt{2\theta}} (\hat{x}-i\hat{y}).
\label{bbd}
\end{eqnarray}
It is easy to verify (using (\ref{xycommutator})) that these operators satisfy the Fock algebra
\beq
[b,b^\dagger ] = 1.
\label{bcomclass}
\eeq
This simply implies that the classical configuration space is isomorphic to boson Fock space,
\beq
\mathcal{H}_c\cong\mathcal{F}\equiv \textnormal{span}\left\{\ket{n} = \frac{(b^\dagger)^n}{\sqrt{n!}}\ket{0}; \;n = 0:\infty  \right\}.
\label{Hclass}
\eeq
Consequently classical configuration space is a Hilbert space, which we shall denote by $\mathcal{H}_c$. This is not an unusual feature, since standard commutative configuration space (i.e., $\Re^2$) is also a Hilbert space. At this point it should be noted that, due to the fact that the non-commutative parameter $\theta$ (which we assume to be of the order of the square of the Planck length (\ref{lp})) is very small, effects of non-commutativity would manifest only at very short length scales, which in turn require very high energies to probe. In this light, it is not sensible to speak of these effects on a classical level, since any uncertainty induced by non-commutativity would manifest on a significantly smaller scale than the uncertainties that are naturally inherent to classical measurements.

As stated, we wish to find a unitary representation of the non-commutative Heisenberg algebra (\ref{heisnc}) on the quantum Hilbert space. It is natural to identify the quantum Hilbert space with the set of Hilbert-Schmidt operators acting on non-commutative configuration space,
\begin{equation}
\label{qhil}
\mathcal{H}_q = \left\{ \psi(\hat{x},\hat{y}): \psi(\hat{x},\hat{y})\in \mathcal{B}\left(\mathcal{H}_c\right),\;\textnormal{tr}_c \left[\psi(\hat{x},\hat{y})^\dagger\psi (\hat{x},\hat{y})\right] < \infty \right\},
\end{equation}
where
\beq
{\rm tr}_c\psi(\hat{x},\hat{y})\equiv \sum_{n=0}^{\infty} \matrel{n}{\psi(\hat{x},\hat{y})}{n}
\eeq
denotes the trace over $\mathcal{H}_c$, and $\mathcal{B}\left(\mathcal{H}_c\right)$ is the set of bounded operators on $\mathcal{H}_c$. In analogy to the Schr\"odinger representation, the square-integrable functions of position co-ordinates are replaced by operators of finite trace (read: norm) which are functions of the position co-ordinates from (\ref{xycommutator}). Of course the physical quantum states of the system are represented by elements of (i.e., operators in) $\hq$. This is indeed a Hilbert space, as demonstrated, for instance in \cite{holevo}. The associated natural inner product on this space is
\begin{equation}\label{inner}
\left(\phi(\hat{x},\hat{y}),\psi(\hat{x},\hat{y})\right) = {\rm tr}_c\left[\phi(\hat{x},\hat{y})^\dagger\psi(\hat{x},\hat{y})\right].
\end{equation}
At this point the clear need for a consistent notation arises, since it is necessary to distinguish between the non-commutative configuration space and the quantum Hilbert space. For this purpose we denote elements of $\mathcal{H}_c$ with angular kets, $|\cdot\rangle$, and  use round kets $\psi(\hat{x},\hat{y})\equiv |\psi)$ for elements of $\hq$. Through the inner product (\ref{inner}) elements of the dual space $\hq^*$ i.e., linear functionals denoted by round bras, $(\psi|$, will map elements of $\hq$ onto complex numbers:
\beq
\left(\phi|\psi\right)=\left(\phi,\psi\right)={\rm tr}_{c}\left[\phi(\hat{x},\hat{y})^\dagger\psi(\hat{x},\hat{y})\right].
\eeq
We distinguish between Hermitian conjugation on $\mathcal{H}_c$ (denoted by $\dagger$) and Hermitian conjugation on $\hq$ (denoted by $\ddagger$). Furthermore, we shall employ capital letters to denote operators acting on $\hq$, whereas lowercase hatted letters are reserved for operators acting on $\mathcal{H}_c$.

With the above framework in place, we are equipped to build the unitary representation of the non-commutative Heisenberg algebra (\ref{heisnc}) on the quantum Hilbert space. This is done in terms of the position operators $X$ and $Y$, and the momentum operators $P_x$ and $P_y$, which act on elements $\psi(\hat{x},\hat{y}) \in \hq$ according to
\begin{eqnarray}
\label{schnc}
{X}\psi(\hat{x},\hat{y}) = \hat{x}\psi(\hat{x},\hat{y})& & \; {Y}\psi(\hat{x},\hat{y}) = \hat{y}\psi(\hat{x},\hat{y})\nonumber\\
{P}_x\psi(\hat{x},\hat{y}) = \frac{\hbar}{\theta}[\hat{y},\psi(\hat{x},\hat{y})] & & \; {P}_y\psi(\hat{x},\hat{y}) = -\frac{\hbar}{\theta}[\hat{x},\psi(\hat{x},\hat{y})].
\end{eqnarray}
Of course this representation conserves the commutation relations of the non-commutative Heisenberg algebra (\ref{heisnc}), and is analogous to the Schr\"{o}dinger representation of the commutative Heisenberg algebra. Position operators act by left multiplication, and momentum operators act adjointly. To make the correspondence more explicit, consider any state $\psi(\hat{x},\hat{y})\in\hq$ that may be written as
\beq
\psi(\hat{x},\hat{y})=\sum_{m,n=0}^\infty c_{m,n} \hat{x}^m\hat{y}^n, \quad c_{m,n} \in \mathbb{C}
\eeq
after suitable ordering. The action of the $x$-momentum operator on this state according to (\ref{schnc}) is simply
\beqa
P_x\psi(\hat{x},\hat{y}) &=& \frac{\hbar}{\theta}[\hat{y},\psi(\hat{x},\hat{y})] \nl
&=& \frac{\hbar}{\theta}(-i\theta)\sum_{m,n=0}^\infty c_{m,n} m\hat{x}^{m-1}\hat{y}^n \nl
&=& -i\hbar \frac{\partial }{\partial \hat{x}}\psi(\hat{x},\hat{y})
\eeqa
Comparing this to (\ref{schrodrep}), the analogy is clear: the momenta from (\ref{schnc}) act as algebraic derivatives.\\

Lastly, we introduce some operators that are simply linear combinations of those in (\ref{schnc}), since these will be convenient to work with in later sections. The first two are linear combinations of the position operators, and represent the counterpart to the operators (\ref{bbd}) on $\hq$:
\begin{eqnarray}
B\equiv\frac{1}{\sqrt{2\theta}}\left({X}+i{Y}\right) &\Rightarrow& B\psi(\hat{x},\hat{y}) = b\psi(\hat{x},\hat{y}),\nonumber\\
B^\ddagger\equiv\frac{1}{\sqrt{2\theta}}\left({X}-i{Y}\right) &\Rightarrow& B^\ddagger\psi(\hat{x},\hat{y}) = b^\dagger\psi(\hat{x},\hat{y}).
\label{BBD}
\end{eqnarray}
The second two are linear combinations of the momentum operators, namely
\beqa
{P}\equiv{P}_x + i{P}_y &\Rightarrow& P\psi(\hat{x},\hat{y})= -i\hbar \sqrt{\frac{2}{\theta}}[b,\psi(\hat{x},\hat{y})], \nonumber\\
{P}^\ddagger \equiv{P}_x -i{P}_y &\Rightarrow& P^\ddagger\psi(\hat{x},\hat{y}) = i\hbar\sqrt{\frac{2}{\theta}}[ b^{\dagger},\psi(\hat{x},\hat{y})].
\label{PPD}
\eeqa
As can be seen from definition (\ref{schnc}) and the commutation relation (\ref{xycommutator}), these two operators commute,
\beq
[P, P^\ddagger] = 0.
\eeq
We require one further notational convention. For any operator $O$ acting on the quantum Hilbert space, we may define left- and right action (denoted by subscripted $L$ and $R$, respectively) as follows:
\beq
O_L\psi=O\psi; \;\;O_R\psi=\psi O \;\;\forall \;\psi \in \mathcal{H}_q.
\label{OLR}
\eeq
In this language, for instance, the complex momenta (\ref{PPD}) may be written as
\beq
P = i\hbar \sqrt{\frac{2}{\theta}}[\br-\bl]\quad \textnormal{and} \quad P^\ddagger = i\hbar \sqrt{\frac{2}{\theta}}[\bld-\brd].
\eeq
Note that left- and right operators always commute, since $O^{(1)}_L O^{(2)}_R\psi=O^{(1)}\psi O^{(2)}= O^{(2)}_R O^{(1)}_L \psi$. Also, if $[O^{(1)}_L,O^{(2)}_L]=c$, then $[O^{(1)}_R,O^{(2)}_R]=-c$. \\ \\

\begin{noindent}
We now have the basic machinery in place to perform calculations in a non-commutative quantum mechanical framework. As far as interpretation is concerned, we proceed essentially as we would with standard quantum mechanics. As will become evident in the section to follow, however, the issue of position measurement must be addressed with great care in our non-commutative framework. This will be done in the context of weak measurements using the language of Positive Operator Valued Measures (POVMs) set out in Section \ref{POVMsection}.
\end{noindent}

\section{Position measurement in non-commutative quantum mechanics: the need for a revised probabilistic framework}
\label{nonlocalposmeas}
As mentioned, the commutation relation (\ref{xycommutator}) induces an uncertainty relation $\Delta \hat x \Delta \hat  y \geq \frac{\theta}{2}$. By implication, it is impossible to measure the co-ordinates $\hat  x$ and $\hat y$ simultaneously to arbitrary accuracy. A corollary to this statement is that it is impossible to define a state that is a simultaneous eigenstate of the operators $X$ and $Y$ in (\ref{schnc}). In commutative quantum mechanics we are able to define such states since this issue does not arise. Clearly the notion of position and its measurement does not exist \emph{a priori} in a non-commutative framework. Again this should be contrasted with position measurements in commutative spaces, which yield complete information about the state of the quantum mechanical system. We shall show here that in order to speak of non-commutative position measurements, it is necessary to invoke some sort of additional structure or missing information, which is manifested in the non-locality of this description.

The natural question to ask is what form the non-commutative analogue to an eigenstate of position would take. We continue by summarising the approach taken in \cite{jopa}. Since $ x$ and $ y$ cannot be specified to arbitrary accuracy in any state, the closest analogue would be a state that has minimal uncertainty on the non-commutative configuration space. Consider the normalised coherent states
\beqa
\ket{z} &=& e^{-z\bar{z}/2}e^{z b^\dagger} \ket{0} \nl
&=& e^{-z\bar{z}/2}\sum_{n=0}^\infty \frac{1}{\sqrt{n!}} z^n \ket{n},
\label{cs}
\eeqa
where $z=\frac{1}{\sqrt{2\theta}}\left(x+iy\right)$ is a dimensionless complex number, and $\zb$ is its complex conjugate. Note that these coherent states are eigenstates of the annihilation operator from (\ref{bbd}),
\beq
b \ket{z} = z\ket{z}, \quad \bra{z}b^\dagger=\zb \bra{z}.
\label{bz}
\eeq
From this and from (\ref{bbd}) we see that
\beqa
\left.
\begin{array}{r}
    \hat{x} = \sqrt{\frac{\theta}{2}}(b+b^\dagger) \\
    \left\langle  \hat{x} \right\rangle =\sqrt{\frac{\theta}{2}} \matrel{z}{b+b^\dagger}{z} =\sqrt{\frac{\theta}{2}} (z+\zb) \\
    \left\langle  \hat{x}^2 \right\rangle ={\frac{\theta}{2}} \matrel{z}{(b+b^\dagger)^2}{z} ={\frac{\theta}{2}} (z^2+\zb^2+2 z \zb + 1) \\
  \end{array}
\right\}
&\Rightarrow & \Delta {\hat x} = \sqrt{\langle \hat{x}^2 \rangle - \langle \hat{x}\rangle^2} = \sqrt{\frac{\theta}{2}}\nl
 \nl
\left.
  \begin{array}{r}
    \hat{y} = i\sqrt{\frac{\theta}{2}}(b^\dagger-b)\\
    \left\langle  \hat{y} \right\rangle = i \sqrt{\frac{\theta}{2}}\matrel{z}{b^\dagger-b}{z} = i\sqrt{\frac{\theta}{2}}(\zb-z) \\
    \left\langle  \hat{y}^2 \right\rangle = {\frac{-\theta}{2}}\matrel{z}{(b^\dagger-b)^2}{z} = {\frac{-\theta}{2}}(z^2+\zb^2- 2z\zb - 1) \\
  \end{array}
\right\}
&\Rightarrow & \Delta {\hat y} = \sqrt{\langle \hat{y}^2 \rangle - \langle \hat{y}\rangle^2} = \sqrt{\frac{\theta}{2}} \nl
\nl
\label{DxDyclass}
\eeqa
Clearly this implies that
\beq
\Delta {\hat x} \Delta {\hat y}=\frac{\theta}{2},
\eeq
i.e., the coherent states (\ref{cs}) display minimum uncertainty in $\hat{x}$ and $\hat{y}$. As is shown in \cite{klauder}, they also admit a resolution of the identity on $\mathcal{H}_c$,
\beqa
\frac{1}{\pi}\int d^2z \ket{z}\bra{z} &=& \frac{1}{\pi}\int d^2z \sum_{n,m=0}^\infty \frac{1}{\sqrt{n!m!}} e^{-z\bar{z}}z^n\zb^m \ket{n}\bra{m} \nl
&=&\frac{1}{\pi} \sum_{n,m=0}^\infty \frac{\ket{n}\bra{m}}{\sqrt{n!m!}}\int_0^{2\pi} d\phi e^{i\phi(n-m)} \int_0^\infty dr \;r e^{-r^2} r^{n+m} \nl
&=& \sum_{n=0}^\infty \ket{n}\bra{n} \nl
&=& \mathbf{1}_c,
\label{identityclass}
\eeqa
where we made use of the Fourier transform representation of the Kronecker delta in polar co-ordinates, and evaluated the radial integral in terms of $\Gamma$-functions. Clearly the states (\ref{cs}) also span $\hc$ since they are simply infinite linear combinations of Fock states. We say that these coherent states provide an over-complete basis on the non-commutative configuration space, where it is important to note that they are not orthogonal,
\beq
\overlap{z_1}{z_2} = e^{-z_1\bar{z}_1/2-z_2\bar{z}_2/2+z_2\bar{z}_1} \neq \delta(z_1-z_2).
\eeq
Turning our attention to the quantum Hilbert space, we construct there a state (operator) that corresponds to (\ref{cs}):
\begin{equation}
\label{braz}
|z )\equiv|z\rangle\langle z|.
\end{equation}
These states are normalised with respect to the inner product (\ref{inner}) and are thus indeed Hilbert-Schmidt operators. Take note, however, of their non-orthogonality,
\beq
\overlapr{z_1}{z_2} = \tr{(\ket{z_1}\bra{z_1})^\ddagger\ket{z_2}\bra{z_2}} = \left| e^{-z_1\bar{z}_1/2-z_2\bar{z}_2/2+z_2\bar{z}_1} \right|^2 = e^{-|z_1-z_2|^2}.
\label{overlapz}
\eeq
Since $z_1$ and $z_2$ are dimensionless here, the Gaussian will become a Dirac delta function in the commutative limit $\theta \rightarrow 0$. From (\ref{BBD}) and (\ref{bz}) it is clear that these states are also eigenstates of $B_L$,
\beq
\bl|z)=z|z).
\eeq
In complete analogy to the calculations in (\ref{DxDyclass}) we can solve for $X$ and $Y$ in (\ref{BBD}), and verify that the states (\ref{braz}) are minimal uncertainty states in position on the quantum Hilbert space:
\beqa
\left.
\begin{array}{r}
    {X} = \sqrt{\frac{\theta}{2}}(\bl+\bld) \\
    \left\langle  {X} \right\rangle_{\ketr{z}} =\sqrt{\frac{\theta}{2}} \matrelr{z}{\bl+\bld}{z} =\sqrt{\frac{\theta}{2}} (z+\zb) \\
    \left\langle  {X}^2 \right\rangle_{\ketr{z}} ={\frac{\theta}{2}} \matrelr{z}{(\bl+\bld)^2}{z} ={\frac{\theta}{2}} (z^2+\zb^2+2 z \zb + 1) \\
  \end{array}
\right\}
&\Rightarrow & \Delta {X} = \sqrt{\langle {X}^2 \rangle - \langle X \rangle^2} = \sqrt{\frac{\theta}{2}}\nl
 \nl
\left.
  \begin{array}{r}
    {Y} = i\sqrt{\frac{\theta}{2}}(\bld-\bl)\\
    \left\langle  {Y} \right\rangle_{\ketr{z}} = i \sqrt{\frac{\theta}{2}}\matrelr{z}{\bld-\bl}{z} = i\sqrt{\frac{\theta}{2}}(\zb-z) \\
    \left\langle  {Y}^2 \right\rangle_{\ketr{z}} = {\frac{-\theta}{2}}\matrelr{z}{(\bld-\bl)^2}{z} = {\frac{-\theta}{2}}(z^2+\zb^2- 2z\zb - 1) \\
  \end{array}
\right\}
&\Rightarrow & \Delta { Y} = \sqrt{\langle {Y}^2 \rangle - \langle Y\rangle^2} = \sqrt{\frac{\theta}{2}} \nl
\nl
\therefore \Delta X \Delta Y = \frac{\theta}{2}. & &
\label{DxDyquant}
\eeqa
It is thus natural to interpret $x$ and $y$ as the dimensionful position co-ordinates. This would imply that the states $\ketr{z}$ are the analogue of position eigenstates on $\hq$, since they saturate the uncertainty relation induced by the commutation relation (\ref{xycommutator}).\footnote{ \label{positionfootnote}In later sections we will show that the states $|z )\equiv|z\rangle\langle z|$ are not the most general states that display the properties discussed above. We follow here, however, the formalism set out in \cite{jopa}, and shall extrapolate on this point in Section \ref{rightsector}.} In this trend, the operator associated with position is $\bl$. It is at this point that we require the probabilistic framework of POVMs set out in Section \ref{POVMsection}. To make use of this formalism, we need to show that the states (\ref{braz}) provide an over-complete set of basis states on the quantum Hilbert space. To prove this, we define the states $\ketr{z,w} \equiv \ket{z}\bra{w}$, and consider that
\beqa
\frac{1}{\pi^2}\int d^2z\int d^2w \;\ketr{z,w} \overlapr{z,w}{\psi}&=& \frac{1}{\pi^2}\int d^2z\int d^2w \; \ketr{z,w}\sum_{n=0}^\infty\matrel{n}{[\ket{z}\bra{w}]^\ddagger\psi}{n} \nl
&=&\frac{1}{\pi^2}\int d^2z\int d^2w \; \ket{z}\bra{w}\matrel{z}{\psi}{w} \nl
&=& \ketr{\psi}.
\eeqa
(In the final step we made use of the fact that $\matrel{z}{\psi}{w}$ is simply a complex number, and of equation (\ref{identityclass})). This implies that
\beq
\frac{1}{\pi^2}\int d^2z\int d^2w \;\ketr{z,w}\brar{z,w}\equiv \mathbf{1}_q
\eeq
is a resolution of the identity on $\hq$. If we now choose $w = z+v$, and note that $d^2 w = d^2 v$ since we are integrating over the entire complex plane, we find that
\beqa
\mathbf{1}_q \ketr{\psi}&=& \frac{1}{\pi^2}\int d^2z\int d^2v \ketr{z,z+v}\overlapr{z,z+v}{\psi} \nl
&=& \frac{1}{\pi^2}\int d^2z\int d^2v \; \ket{z}\bra{z+v}\matrel{z}{\psi}{z+v} \nl
&=& \frac{1}{\pi^2}\int d^2z\int d^2v \; e^{-|v|^2} \ket{z}\bra{z}e^{\vb\stackrel{\leftarrow}{\partial_{\bar{z}}}+v\stackrel{\rightarrow}{\partial_z}}\matrel{z}{\psi}{z} \nl
&=& \frac{1}{\pi}\,\int d^2z \;\ketr{z}e^{\stackrel{\leftarrow}{\partial_{\bar{z}}}\stackrel{\rightarrow}{\partial_z}}\overlapr{z}{\psi},
\label{identityzproof}
\eeqa
where we have defined $\partial_{\bar z}\equiv \frac{\partial}{\partial \bar{z}}$ and $\partial_z\equiv \frac{\partial}{\partial z}$, used the fact that $e^{v{\partial_z}}f(z) = f(z+v)$, and performed the Gaussian integral over $v$ explicitly. Consequently
\begin{equation}
\label{identityz}
\mathbf{1}_q=\frac{1} {\pi}\int d^2z \; |z)e^{\stackrel{\leftarrow}{\partial_{\bar{z}}}\stackrel{\rightarrow}{\partial_z}}(z|\equiv\frac{1} {\pi}\int d^2z \; |z)\star\brar{z}
\end{equation}
is a resolution of the identity on $\hq$, and it follows that the operators

\begin{equation}
\label{povm}
\pi_z=\frac{1}{\pi}|z) e^{\stackrel{\leftarrow}{\partial_{\bar{z}}}\stackrel{\rightarrow}{\partial_z}}(z|\;,\quad \int d^2z \;\pi_z=\mathbf{1}_q\,
\end{equation}
provide an Operator Valued Measure. (Note that since $x$ and $y$ are dimensionful co-ordinates, we have that $d^2z = \frac{dxdy}{2\theta}$). The operators are also positive, since
\beqa
\matrelr{\phi}{\pi_z}{\phi} &=&\frac{1}{\pi} \overlapr{\psi}{z}\star\overlapr{z}{\psi} \nl
&=& \frac{1}{\pi}\sum_{n=0}^\infty \frac{1}{n!} \frac{\partial^n\overlapr{\phi}{z}}{\partial\zb^n}\frac{\partial^n\overlapr{z}{\phi}}{\partial z^n} \nl
&=& \frac{1}{\pi}\sum_{n=0}^\infty \frac{1}{n!}\left|\frac{\partial^n\overlapr{z}{\phi}}{\partial z^n}\right|^2 \geq 0 \;\; \forall \;\phi.
\eeqa
Assuming that the system is in a pure state $\ketr{\psi}$, it is thus consistent to assign the probability of finding the particle at position $\left(x,y\right)$ (defined in terms of $z$ and $\zb$) as
\begin{eqnarray}
P(z,\zb)&=&\left(\psi|\pi_z|\psi\right)\nl
&=& \frac{1}{\pi}\overlapr{\psi}{z}\star\overlapr{z}{\psi}.
\label{Pz}
\end{eqnarray}
Consider the difference of this position probability distribution to those in standard (commutative) quantum mechanics. In the standard case, we simply define the probability distribution as the modulus squared of the position wave function: $P(\vec{x})\equiv|\psi(\vec{x})|^2$. This is not the case here --- a star product is involved. Consequently, if we define $\psi(z,\zb)\equiv\overlapr{z}{\psi}$ as the wave function in position, we must always bear in mind that it is not a probability amplitude in the standard sense, but that the star product is required to form the probability distribution. It is in this sense that this probabilistic framework is non-local, since we require knowledge of all orders of derivatives of the overlap $\overlapr{z}{\psi}$. The overlap $\psi(z,\zb)\equiv\overlapr{z}{\psi}$ does thus not provide complete information about the state $\ketr{\psi}$.

Lastly, we shall address the post-measurement state of the system. In Appendix \ref{POVMsquaredProof} we show that for an element $\pi_z$ of the POVM (\ref{povm}) we have the property that
\beq
\pi_z^{1/2} = \sqrt{\pi}\pi_z.
\label{rootpiz}
\eeq
We conclude that the elements of the POVM (\ref{povm}) are (up to a constant) simply projectors.\footnote{ Note that this does not imply that we do not need to use the language of POVMs here, however, since it is clear that separate elements of the POVM are not orthogonal, $\pi_z \pi_w \neq \delta(z-w) \pi_z$. } This allows us to construct the post-measurement state by considering the discussion from Section \ref{newpostulates}. We recall that the form of the detection operators was simply the square root of the corresponding POVM elements (up to a unitary transformation), or in this particular case
\beq
D_z = U \pi_z^{1/2}=\sqrt{\pi}U\pi_z.
\eeq
The state of the system initially in pure state $\ketr{\psi}$ after measurement is now simply
\beq
\ketr{\phi} = \frac{D_z \ketr{\psi}}{\sqrt{\matrelr{\psi}{D_{z}^\dagger D_z}{\psi}}}=\frac{\sqrt{\pi}U\pi_z \ketr{\psi}}{\pi\sqrt{\matrelr{\psi}{\pi_z^2}{\psi}}}=\sqrt{\pi}\frac{U\pi_z \ketr{\psi}}{\sqrt{\matrelr{\psi}{\pi_z}{\psi}}}.
\eeq
Again, it is important to note that the unitary transformation above reflects that we do not have complete information about the post-measurement state.\\

\hbox{}
This completes our review of the non-commutative quantum mechanical formalism set out in \cite{jopa}. In further parts of this thesis we shall consider in greater detail the POVM (\ref{povm}). As already alluded to in footnote \ref{positionfootnote}, the states (\ref{braz}) are not the only ones in $\hq$ that display minimal uncertainty in $x$ and $y$. It is also clear that the description of position measurements in the framework above is highly non-local, in that it requires the knowledge of all orders of derivatives in $z$ and $\zb$ of the wave function $\psi(z,\zb)\equiv\overlapr{z}{\psi}$. We shall show in Chapter \ref{rightsectorchapter} that the introduction of additional degrees of freedom (that characterise the right sector of basis states) allows us to decompose the resolution of the identity (\ref{identityz}) in such a way that the resulting POVM is local in $z$, thereby allowing for local descriptions of position measurements. This, of course, brings with it several interpretational questions regarding the meaning of the added degrees of freedom. Focusing on two particular choices of bases, we shall attempt to provide some insight into possible physical interpretations of the right sector in Chapters \ref{zvchapter}, \ref{znchapter}. In Chapter \ref{gaugechapter} we shall demonstrate that the right sector degrees of freedom may also be thought of as gauge degrees of freedom in a gauge-invariant formulation of non-commutative quantum mechanics. 
\chapter{THE RIGHT SECTOR AND BASES FOR LOCAL POSITION MEASUREMENTS}
\label{rightsectorchapter}
As alluded to in the previous chapter, the motivation for introducing the states $\ketr{z}\equiv \ket{z}\bra{z}$ was that they are optimally (minimally) localised in $\hq$, in the sense that they saturate the $x-y$ uncertainty relation. In this regard they may be considered as being the analogue of position eigenstates (of the associated non-Hermitian operator $\bl$) on the quantum Hilbert space. In this chapter we shall show that these states are not the most general states to display these properties. The description of position measurements in terms of these states is, as stated, non-local in that it requires the knowledge of all orders of derivatives in $z$ and $\zb$ of the overlap $\psi(z,\zb)\equiv\overlapr{z}{\psi}$. By introducing new degrees of freedom that contain explicit information about the right sector of quantum states, we shall now find decompositions of the identity (\ref{identityz}) on $\hq$ in terms of bases that allow a local description of position measurements. (By ``local'' we mean that these descriptions do not require explicit knowledge of all orders of derivatives in $z$ and $\zb$). Naturally we wish to attach physical meaning to these newly introduced degrees of freedom. This matter will be addressed in later chapters where, for instance, we shall consider arguments from the corresponding classical theories which indicate that the notion of additional structure is clearly encoded in these variables.

\section{Arbitrariness of the right sector in non-local position measurements}
\label{rightsector}

Let us revisit some arguments from Chapter \ref{ncformalism}. In equation (\ref{DxDyquant}) we showed that the states $\ketr{z}$ are minimally localised in the variables $x$ and $y$, and are eigenstates of the operator that we associate with position, $\bl$. Consider now the state
\beq
\ketr{z,\phi}\equiv \ket{z}\bra{\phi}\;\in\hq,
\label{zphi}
\eeq
where $\phi$ is arbitrary. Next we note that, for instance,
\beqa
\left\langle  {X} \right\rangle_{\ketr{z,\phi}}&=&\sqrt{\frac{\theta}{2}} \matrelr{z,\phi}{\bl+\bld}{z,\phi}\nl
&=& \sqrt{\frac{\theta}{2}} \tr {[\ket{z}\bra{\phi}]^\ddagger (\bl+\bld) \ket{z}\bra{\phi}}\nl
&=& \sqrt{\frac{\theta}{2}}\matrel{z}{b+b^\dagger}{z}\nl
&=&\sqrt{\frac{\theta}{2}} (z+\zb).
\eeqa
Thus it is clear that all expectation values taken with respect to the basis elements $\ketr{z}$ in (\ref{DxDyquant}) are independent of the right sector $\bra{\cdot}$ of these elements, since the trace of the inner product on $\hq$ essentially removes this information (see the second and third line in the calculation above). Consequently the expectation values of the same (left-acting) operators with respect to the basis elements $\ketr{z,\phi}$ equal those taken with respect to $\ketr{z}$. Following the same arguments as previously, we thus note that the states (\ref{zphi}) are also minimum uncertainty states in position \emph{for all} $\phi$, i.e., this statement is independent of the specific form of the right sector. Furthermore, these states are also eigenstates of our position operator,
\beq
\bl \ketr{z,\phi} = b \ket{z}\bra{\phi} = z \ket{z}\bra{\phi} \;\; \forall\; \phi.
\eeq
We conclude that the non-local framework for position measurements from \cite{jopa} set out in Section \ref{nonlocalposmeas} is insensitive to information contained in the right sector of states of the form $\ket{z}\bra{\cdot}$. Consider the contrast to a 2 dimensional commutative quantum system, where states can be completely specified by knowledge of position, i.e., $x$ and $y$ (since the corresponding observables form a maximally commuting set). In lieu of the above arguments, however, it becomes clear that in the non-commutative framework additional information from the right sector is necessary to specify states completely. Since the non-locality in position measurements as in Section \ref{nonlocalposmeas} is a direct consequence of the star product, one may ask whether a manifestly local description in terms of a decomposition of this star product is possible. We shall address this question below.

\section{Decomposition of the identity on $\hq$}
\label{zalphasection1}
The first requirement for a probability description in terms of POVMs is a resolution of the identity on the quantum Hilbert space. For this purpose, suppose we have a set $\left\{ \ket{\alpha} \right\}$ of states in $\hc$ that satisfy
\beq
O_R (\ket \phi \bra{\alpha}) = \ket \phi \bra \alpha O = \lambda_\alpha \ket \phi \bra{\alpha}\;\forall \;\ket \phi \in \hc, \; \sum_\alpha \ket{\alpha}\bra{\alpha} = \mathbf{1}_c.
\label{alpha}
\eeq
(Here the state label $\alpha$ could be discrete or continuous. In the latter case, the summation would simply be replaced by an integral.) The aim is to find a decomposition of the identity (\ref{identityz}) on $\hq$ in terms of states $\ketr{z,\alpha}$ where $\alpha$ specifies the right sector of an outer product as in (\ref{zphi}). If we achieve this, we have a new set of states that are still eigenstates of $\bl$, but that have an added state label (degree of freedom). Such a state is of course a minimum uncertainty state in $x$ and $y$ (as discussed in \ref{rightsector}) that is localised at $z$. If we had a transformation that would localise the state elsewhere (i.e., translate $z$), we would require that this transformation is unitary and maintains the minimum uncertainty property of the states. To this end, let us define the operator
\beqa
T\left(z\right)&\equiv&e^{-\frac{i}{\hbar}\sqrt{\frac{\theta}{2}}(\bar{z}P+zP^\ddagger)}\nl
&=&e^{z(\bld-\br)+\zb(\br-\bl)},
\label{Tz}
\eeqa
which acts on any $\phi \in \hq$ according to\footnote{ To show this we use definition (\ref{PPD}) of the complex momenta.}
\beq
T(z)\phi = e^{z b^\dagger-\zb b}\,\phi\, e^{\zb b-z b^\dagger},
\label{Tzphi}
\eeq
and is unitary with respect to the inner product (\ref{inner}). As with usual translations, we have that
\beq
T(z)T^\ddagger(w) = T(z-w), \quad \textnormal {and} \quad \left[ T(z),T(w) \right] = 0.
\label{Tzproperties}
\eeq
Though this operator is the direct analogue of the translation operator $e^{-\frac{i}{\hbar} \vec{p}\cdot\vec{x}}$ from standard quantum mechanics\footnote{ To see this, take note that for two complex variables $u=u_x+iu_y$ and $v=v_x+iv_y$ we have that $(u\bar v + \bar u v)/2 = u_x v_x + v_x u_x$ is simply the dot product. Applying this to the definitions (\ref{BBD}) and (\ref{PPD}) makes the analogy clear.}, take note of its left \emph{and} right action. In this light it is clear that the state (\ref{braz}) may be written as $\ketr{z}=T(z)\ket{0}\bra{0}$, as is seen by splitting the exponents in (\ref{Tz}) through the identity $e^{A+B}=e^A e^B e^{-1/2[A,B]}$ which applies whenever the operators $A$ and $B$ commute to a constant. It would thus make sense to introduce states of the form
\beq
\ketr{z,\alpha}\equiv T(z) \left(\ket{0}\bra{\alpha}\right)=\ket{z}\bra{\alpha}e^{\zb b-z b^\dagger},
\label{Tzalpha}
\eeq
which simply represent some state $\ket 0 \bra \alpha$ that was originally located at the origin, and was then translated to the point $z$. Returning to (\ref{alpha}), we note that these states are eigenstates of the \emph{translated} operator $O_R$,
\beq
T(z)O_RT^\ddagger(z) \ketr {z,\alpha} = T(z)\left( \ket 0 \bra \alpha O\right) = {\lambda}_\alpha \ketr {z,\alpha}.
\eeq
This makes sense since we have translated the state $\ket 0 \bra \alpha$ away from the origin to the point $z$, and consequently we would expect to have to shift the operator $O_R$ to this point in order to satisfy the eigenvalue equation from (\ref{alpha}). Clearly a translated state of the form (\ref{Tzalpha}) above is still an eigenstate of $\bl$ and also a minimum uncertainty state (since the arguments from Section \ref{rightsector} hold also for the states (\ref{Tzalpha})).\\ \\

To show that a resolution of the identity on $\hq$ in terms of these states is possible, consider that
\beqa
\frac{1}{\pi}\int d^2z \sum_\alpha \overlapr{\psi}{z,\alpha} \overlapr{z,\alpha}{\phi} &=& \frac{1}{\pi}\int d^2z \sum_\alpha \tr{\psi^\ddagger\ket{z}\bra{\alpha}e^{\zb b-z b^\dagger}} \tr{e^{z b^\dagger-\zb b}\ket{\alpha} \bra z \phi} \nl
&=&\frac{1}{\pi}\int d^2z \sum_\alpha \bra \alpha e^{\zb b-z b^\dagger} \psi^\ddagger \ket{z} \bra z \phi e^{z b^\dagger-\zb b} \ket \alpha \nl
&=& \frac{1}{\pi}\int d^2z \bra z \phi \psi^\ddagger \ket z \nl
&=& \tr{\phi \psi^\ddagger} \nl
&=& \tr{\psi^\ddagger \phi} \nl
&=& \overlapr{\psi} {\phi},
\label{identityzalphaproof}
\eeqa
where we made use of the completeness relation (\ref{alpha}), the definition of the trace over $\hc$ in terms of the classical coherent states (\ref{cs}) and the cyclic property of the trace. We have thus shown that
\beq
\frac{1}{\pi}\int d^2z \sum_\alpha \ketr{z,\alpha}\brar{z,\alpha} = \mathbf 1 _q
\label{identityzalpha}
\eeq
is a resolution of the identity on $\hq$ for any set $\{\ket \alpha\}$ of states in $\hc$ that satisfies (\ref{alpha}). This implies that
\beq
|z)\star\brar{z} = \sum_\alpha \ketr{z,\alpha}\brar{z,\alpha},
\eeq
i.e., that we have decomposed the star product in terms of a new variable $\alpha$ which characterises the right sector of the states (\ref{Tzalpha}). This procedure simply reflects that the ``missing information'' encoded in the non-local description set out in Chapter \ref{ncformalism} may be made explicit through the introduction of new degrees of freedom. This makes manifest the additional structure that was alluded to earlier.

\section{POVMs for local position measurements}
\label{zalphasection2}

We depart by noting that the states $\ketr {z,\alpha}$ not only admit a resolution of the identity on $\hq$, but also that the corresponding operators
\beq
\pi_{z,\alpha}\equiv\frac{1}{\pi}\ketr {z,\alpha} \brar {z,\alpha}
\eeq
are positive and Hermitian (this is easy to see --- consider equation (\ref{Pzalpha})). Thus we have a POVM,
\beq
\int d^2z \sum_\alpha \pi_{z,\alpha} = \mathbf 1 _q,\quad \pi_{z,\alpha} \geq 0,
\eeq
in terms of which we can ask probabilistic questions according to Section \ref{POVMsection}. We could, for instance, ask what the probability distribution in $z$ and $\alpha$ is, given that the system is in a pure state $\ketr \psi$. This is simply
\beq
P(z,\alpha) = \matrelr{\psi}{\pi_{z,\alpha}}{\psi}= \frac{1}{\pi}\overlapr{\psi}{z,\alpha}\overlapr{z,\alpha}{\psi}=\frac{1}{\pi}| \overlapr{z,\alpha}{\psi} |^2.
\label{Pzalpha}
\eeq
This distribution provides information not only about position, but also about the degree of freedom $\alpha$. It also stands in contrast to the probability distribution (\ref{Pz}) in $z$ , in that $\overlapr{z,\alpha}{\psi}$ is indeed a probability amplitude in the standard sense: its modulus squared is the probability distribution, and there is no need for a star product. Also note that since
\beq
\sum_\alpha \pi_{z,\alpha}=\frac{1}{\pi}\ketr z \star \brar z = \pi_z,
\eeq
where $\pi_z$ refers to the POVM (\ref{povm}), we may obtain the probability distribution (\ref{Pz}) by summing (\ref{Pzalpha}) over all $\alpha$. Similarly, we could obtain a distribution in $\alpha$ only by integrating (\ref{Pzalpha}) over $(z,\zb)$.

Let us take stock of the discussion thus far. We have decomposed the star product by introducing a new degree of freedom which characterises the right sector of the resulting states. This allows us to write a probability distribution in position and in this new variable --- a distribution that is manifestly local in $z$ and $\zb$, in that it does not require knowledge of all orders of derivatives in these variables. Summation over all possible values of this new degree of freedom returns us to the non-local description in position only, where we do require explicit knowledge of said derivatives. The price to pay for the convenience of the local description with the added degree of freedom is that we are as yet unsure of the physical meaning of the new degree of freedom. What is clear, however, is that this description resolves more transparently the information that is encoded through derivatives in the non-local description. One should note, however, that the two description contain the \emph{same} information --- it is simply accessed in different ways.

A further use of the completeness relations (\ref{identityz}) and (\ref{identityzalpha}) is that we may reconstruct any state $\ketr{\phi}$ from overlaps of the form $\left(z|\phi\right)$ and $\left(z,\alpha|\phi\right)$. As stated, the former results in a non-local description, whereas the latter is local in position; both descriptions address the same physical information, and one does not display a loss of information when compared to the other. We thus have two types of bases that may be used to represent physical systems --- one non-local and the other local in position. As is to be expected, the local basis is mathematically more convenient to work with since we need not access higher order derivatives. We shall demonstrate later, however, that constraints may arise in the local description. These constraints must be handled with caution, and restrict which states in the system are physical.

It should be noted from the onset that the additional degrees of freedom in our local descriptions differ fundamentally from additional quantum labels (such as spin) from standard quantum mechanics: in the standard setting such quantum labels must be added in by hand, whereas these additional state labels appear naturally and unavoidably in any local position description of non-commutative quantum mechanics. For the remainder of this thesis we shall concern ourselves with two particular choices of bases that allow such local probability descriptions --- one with a continuous state label for the right sector, and the other with a discrete label. We shall attempt to explore the physical meaning of the additional degrees of freedom in each case, and represent a few non-commutative quantum systems in these bases in order to gain understanding of the additionally resolved information. Thereafter we shall demonstrate that local transformations between bases for the right sector may also be thought of as gauge transformations in a gauge-invariant formulation of the theory.

\chapter{THE BASIS $\ketr{z,v}\equiv T(z)\ket 0 \bra v $}
\label{zvchapter}

In this chapter we shall introduce a basis of the form (\ref{Tzalpha}) where the right sector is characterised by a coherent state with label $v$. After showing that a resolution of the identity on $\hq$ in terms of these states is possible, we shall construct the positive, Hermitian elements of the associated POVM, thereby providing a probability formulation in terms of this basis. Thereafter we consider the associated classical theories to gain insights into the physical nature of the degree of freedom $v$, and apply the basis to representing a few non-commutative quantum mechanical problems that were investigated in \cite{jopa}.

\section{Decomposition of the identity on $\hq$ and the associated POVM}
\label{zvsection1}
Let us take a look at the derivation of the identity on the quantum Hilbert space. From equation (\ref{identityzproof}) it is clear that the star product may be written as
\beq
\star = e^{\stackrel{\leftarrow}{\partial_{\bar{z}}}\stackrel{\rightarrow}{\partial_z}}=\int d^2v \; e^{-|v|^2} e^{\vb\stackrel{\leftarrow}{\partial_{\bar{z}}}+v\stackrel{\rightarrow}{\partial_z}}.
\eeq
If we now introduce the states
\beq
\ketr{z,v} \equiv e^{-v\vb/2} e^{\vb\partial_{\zb}}\ketr{z},
\label{zv1}
\eeq
we note that the identity on the quantum Hilbert space may be written as
\beq
\mathbf{1}_q=\frac{1}{\pi^2} \int d^2z \int d^2v\ketr{z,v}\brar{z,v}.
\label{identityzv}
\eeq
Considering the states (\ref{zv1}) in more detail, we observe that
\beqa
\ketr{z,v} &=& e^{-v\vb/2} e^{\vb\partial_{\zb}}\ketr{z} \nl
&=& T(z)\ket{0}\bra{v} \nl
&=& e^{\frac{1}{2}(\bar{z}v-\bar{v}z)}\ket{z}\bra{z+v},\,\textnormal{with} \; z,v \in \mathbb{C}.
\label{zv2}
\eeqa
Here $T(z)$ denotes the translation operator (\ref{Tz}). It is also evident that these states may be viewed as ``position eigenstates'' in the sense of Section \ref{rightsector},
\beq
\bl \ketr{z,v} =e^{\frac{1}{2}(\bar{z}v-\bar{v}z)} \, b\ket{z}\bra{z+v} = z\ketr{z,v}.
\eeq
This statement, in fact, holds even for linear combinations of these states taken over $v$. In addition to the resolution of the identity on $\hq$ in terms of the states $\ketr{z,v}$, we also have the required positivity condition,
\beqa
\overlapr{\phi}{z,v}\overlapr{z,v}{\phi} &=& e^{-|v|^2} \overlapr{\psi}{z}e^{\vb\stackrel{\leftarrow}{\partial_{\bar{z}}}+v\stackrel{\rightarrow}{\partial_z}}\overlapr{z}{\psi} \nl
&=& e^{-|v|^2} \left[ e^{\vb\partial_{\zb}}\overlapr{\psi}{z}\right] \left[e^{v\partial_z}\overlapr{z}{\psi} \right]\nl
&=& e^{-|v|^2} \left|e^{v\partial_z}\overlapr{z}{\psi}\right|^2 \geq 0.
\eeqa
Thus we have a new POVM, namely
\beq
\pi_{z,v}\equiv\frac{1}{\pi^2}\ketr{z,v}\brar{z,v}, \quad \int d^2z \int d^2v \; \; \pi_{z,v} = \mathbf{1}_q.
\label{pizv}
\eeq
Correspondingly, we may define a probability distribution in $z$ and $v$. Assuming that the system is in a pure state $\ketr{\psi}$, this is simply
\beq
\label{Pzv}
P(z,v)=\brar{\psi}\pi_{z,v}\ketr{\psi}=\frac{1}{\pi^2}\overlapr{\psi}{z,v}\overlapr{z,v}{\psi} =\frac{1}{\pi^2} \left| \overlapr{z,v}{\psi} \right|^2.
\eeq
This probability provides information not only about position $z$, as was the case in (\ref{Pz}), but also about a further degree of freedom, $v$. As stated, the two distributions are connected, in that we could also ask for the probability to find the particle localised at point $z$, without detecting any information regarding $v$. This is simply the sum of the probabilities (\ref{Pzv}) over all $v$:
\beq
\label{probt}
P(z)=\frac{1}{\pi}\int d^2v\,\, P(z,v)=\brar{\psi}\left[\frac{1}{\pi}\int d^2v\,\,\pi_{z,v}\right]\ketr{\psi}=\brar{\psi}\pi_{z}\ketr{\psi},
\eeq
with $\pi_z$ as in (\ref{povm}).

\hbox{}

To summarise, we have found states that allow a decomposition of the identity (\ref{identityz}) on $\hq$ through the introduction of added degrees of freedom $v$ which characterise the right sector of the state in terms of a coherent state. As set out in Sections \ref{rightsector} and \ref{zalphasection1}, these states are position states. Since the relevant positivity criteria are met, we are able to construct a POVM in terms of these states (as in Section \ref{zalphasection2}), which can be used to ask local probabilistic questions. Of course the price to pay for this local description is that it is unclear what the physical meaning of the newly introduced degree of freedom $v$ is. This matter will be addressed in the remainder of this chapter.

\section{An analysis of the corresponding classical theory}
\label{class}

As stated, the states $|z,v)$ form an over-complete coherent state basis for the quantum Hilbert space of the non-commutative system. Consequently we may derive a path integral action in the standard way according to \cite{klauder} (this calculation is done explicitly in Appendix \ref{PathIntAct}). This action is generally given by
\beq
S=\int_{t'}^{t''}dt\, \matrelr{z,v}{i\hbar \frac{d}{dt} - H}{z,v},
\label{classact1}
\eeq
where we take the states $\ketr{z,v} \equiv \ketr{z[t],v[t]}$ to be time-dependent. We consider here a non-commutative Hamiltonian of the form $H=\frac{P^2}{2m}+V(X,Y)$. In order to compute this action explicitly we thus require the diagonal matrix elements in the $|z,v)$ basis of the time-derivative operator and of the kinetic and potential terms of the Hamiltonian. Note that since $\ketr{z,v}=e^{-(z\zb+\vb z+ v \vb/2)}e^{z b^\dagger}\ket 0 \bra 0 e^{(\zb+\vb)b}$, we have
\beqa
\matrelr{z,v}{\frac{d}{dt}}{z,v}&=& \matrelr{z,v}{-(z\dot \zb +\dot z \zb+ \dot\vb z+ \dot z \vb + [v\dot \vb+ \dot v \vb]/2)+\dot z \bld + (\dot \zb \dot \vb)\br}{z,v} \nl
&=& \dot \zb v - \vb\dot z + \frac{1}{2}(\dot \vb v - \dot v \vb).
\label{classact2}
\eeqa
The free part of the Hamiltonian is simply $\frac{1}{2m}PP^\ddagger$. Through (\ref{PPD}) we obtain
\beqa
\matrelr{z,v}{PP^\ddagger}{z,v}&=&\matrelr{z,v}{-\hbar^2(2/\theta)[\br-\bl][\bld-\brd]}{z,v}\nl
&=& \frac{2\hbar^2}{\theta}\vb v.
\label{classact3}
\eeqa
Since this term represents the kinetic energy, we see that $v$ has a clear connection to momentum in this context (namely that it equals [up to constants] the expectation value thereof in the basis (\ref{zv2})). Lastly, the potential may be written as a normal ordered function of $\bl$ and $\bld$ by solving $X$ and $Y$ in (\ref{BBD}), and thus its matrix element is simply
\beq
\matrelr{z,v}{V(X,Y)}{z,v}=\matrelr{z,v}{V(\bld,\bl)}{z,v}=V(\zb,z),
\label{classact4}
\eeq
where it is important to note that this potential does not depend on $v$. Inserting (\ref{classact2}), (\ref{classact3}) and (\ref{classact4}) into (\ref{classact1}), we obtain
\beq
\label{action1}
S=\int_{t'}^{t''} dt\left[i\hbar\left(\dot \zb v - \vb\dot z + \frac{1}{2}(\dot \vb v - \dot v \vb)\right)-\frac{\hbar^2}{m\theta}\bar{v}v-V(\bar z,z)\right].
\eeq

In order to gain some physical intuition about this system, we proceed to show that this action can be identified precisely with that of \cite{susskind} in the case of a free Hamiltonian, i.e., when $V=0$. This picture makes the notion of extent and structure very explicit, in that it entails two particles of mass $m$ and opposite charge $\pm q$ moving in a magnetic field $\vec B = B \hat z$ perpendicular to the plane. The charges further interact through a harmonic interaction.\footnote{See Figure \ref{fig:chargesimage} for a schematic.} If we assign  $z$ to be the dimensionful co-ordinates of one particle and $z+v$ the dimensionful co-ordinates of the other (i.e., $v$ is the relative co-ordinate), we observe that the Lagrangian of this system in the symmetric gauge and in S.I. units is
\beq
L=\frac{1}{2}m\dot{\bar z}\dot{z}+\frac{1}{2}m\left(\dot{\bar z}+\dot{\bar v}\right)\left(\dot{z}+\dot{v}\right)+\frac{iqB}{4c}\left(\dot{\bar z}z-\bar z\dot z\right)-\frac{iqB}{4c}\left[\left(\dot{\bar z}+\dot{\bar v}\right)\left(z+v\right)-\left({\bar z}+{\bar v}\right)\left(\dot{z}+\dot{v}\right)\right]-\frac{1}{2}K\bar{v}{v}.
\eeq
Here the first two terms are the kinetic energy terms, the third and fourth terms represent the coupling to the magnetic field and the last term is the harmonic potential with spring constant $K$. Introducing the magnetic length $\ell=\sqrt{\frac{2\hbar c}{qB}}$ and the dimensionless co-ordinates $\frac{z}{\ell}$ and $\frac{v}{\ell}$ this reduces to
\beq
L=\frac{1}{2}m\ell^2\dot{\bar z}\dot{z}+\frac{1}{2}m\ell^2\left(\dot{\bar z}+\dot{\bar v}\right)\left(\dot{z}+\dot{v}\right)+i\hbar\left[\left(\dot{\bar z}v-\bar v\dot z\right)+\frac{1}{2}\left(\dot{\bar v}v-{\bar v}\dot v\right)\right]-\frac{1}{2}K\ell^2\bar{v}{v}.
\eeq
In the limit of a strong magnetic field where $\ell\rightarrow 0$ the kinetic terms may be ignored. In this case this Lagrangian reduces to that in (\ref{action1}), where we identify $K=\frac{2\hbar^2}{m\ell^2\theta}$. Given the physical picture described here, it is clear that in this context $v$ clearly represents the spatial extent of this two-charge composite. Note that in the strong magnetic field limit the spring constant becomes very large. The physical consequence of this is that internal mode excitations are suppressed, and the composite behaves more like a stiff rod whose length is proportional to its (average) momentum (see (\ref{classact3}) and the subsequent observation).

Let us now return to (\ref{action1}) for the case where the potential is non-zero. As stated, the potential may be written as a function of $z$ and $\zb$ through appropriate normal ordering, and is independent of $v$. One should note, however, that the normal ordering would generate $\theta$-dependent corrections, i.e., it is not simply the naive potential obtained by replacing the non-commutative variables with commutative ones. In this sense it is different from the classical potential of a point particle to which it reduces in the commutative limit. In \cite{sunandan} a non-local form of the path integral action was found. This action is later cast into a manifestly local form through the introduction of auxiliary fields. Comparing equation (13) of said article to equation (\ref{action1}), it is immediately evident that the variable $v$ plays exactly the same role as the auxiliary fields --- non-locality is remedied through the introduction of added degrees of freedom. The properties of this action were already discussed there;  in particular it was found that this is a second class constrained system that yields, upon introduction of Dirac brackets, non-commuting co-ordinates $z$ and $\bar z$ as one would expect. We shall point out in the next section that constraints also arise on the quantum mechanical level.

Continuing, we note that through use of the Lagrangian
\beq
L=\left[i\hbar\left(\dot \zb v - \vb\dot z + \frac{1}{2}(\dot \vb v - \dot v \vb)\right)-\frac{\hbar^2}{m\theta}\bar{v}v-V(\bar z,z)\right]
\label{classLagrangian}
\eeq
from (\ref{action1}) it is easy to obtain the equations of motion through the Euler-Lagrange equations $\frac{d}{dt}\frac{\partial L}{\partial \dot q} - \frac{\partial L}{\partial q} = 0$ with $q\in\{ z,\zb,v,\vb \}$. These simply read
\beqa
\label{eqmot1}
i\hbar \dot{\bar v}-\frac{\partial V}{\partial z}&=&0,\nonumber\\
-i\hbar \dot{v}-\frac{\partial V}{\partial \bar z}&=&0,\nonumber\\
i\hbar \left(\dot{\bar z}+\dot{\bar v}\right)-\frac{\hbar^2}{m\theta}\bar v&=&0,\nonumber\\
-i\hbar \left(\dot{z}+\dot{v}\right)-\frac{\hbar^2}{m\theta}v&=&0.
\eeqa
Solving for $v$ and $\vb$ in the third and fourth lines above, inserting this into the first two equations and finally reintroducing the dimensionful variable $z\rightarrow \frac{z}{\sqrt{2\theta}}$, this can be cast into a more recognisable form,
\beqa
\label{eqmot2}
\ddot z=-\frac{2}{m} \frac{\partial V}{\partial \bar z}-\sqrt{2\theta} \ddot v,\nonumber\\
\ddot{\bar z}=-\frac{2}{m} \frac{\partial V}{\partial z}-\sqrt{2\theta} \ddot{\bar v}.
\eeqa
(The factor of 2 in the first term is indeed correct since $\partial_z=\frac{1}{2}[\partial_x-i\partial_y]$). We observe that, up to leading order in $\theta$, the position obeys the standard equations of motion. The additional terms reflect the coupling to the variable $v$. This coupling is observed also on the quantum mechanical level, as will be seen for instance in the case of the harmonic oscillator in Section \ref{harmos}.

The conserved energy is given in terms of the dimensionful variable $z$ and dimensionless variable $v$ as
\beq
\label{en1}
E=\frac{\hbar^2}{m\theta}\bar{v}v+V(z,\zb),
\eeq
where we computed the time derivative $\frac{dE}{dt}$ explicitly and used (\ref{eqmot1}). From this it is clear that the momentum canonically conjugate to $z$ is $-i\hbar\bar v$. This again reflects the direct relation between $v$, which we associate with spatial extent, and momentum. We observe that, as reflected in (\ref{eqmot1}), the momentum conjugate to $z$ is not simply $m\dot z$ as would be the case for a point particle. This in turn signals that the conserved energy is not just the sum of kinetic and potential energies of a point particle. To illustrate this explicitly, we rewrite (\ref{en1}) as
\beq
E=\frac{m}{2}\dot{\bar z}\dot{z}+V-m\theta\dot{\bar v}\dot{v}+i\hbar\left(v\dot{\bar v}-\dot{v}\bar{v}\right),
\eeq
where we again made use of (\ref{eqmot1}). From (\ref{eqmot2}) we see that the dimensionful $z$ has a length scale $\ell_z$, determined by the potential, associated with it. Using this in the first two equations of (\ref{eqmot1}), we conclude that the dimensionless $v\sim \frac{\sqrt\theta}{\ell_z}$, which implies the vanishing of the correction terms in the commutative limit. It is, of course, natural that the particular dynamics of a system would govern the positional length scales involved. This generic phenomenon is also demonstrated explicitly on the quantum mechanical framework in Section \ref{harmos} in the context of the harmonic oscillator.

Lastly we note from (\ref{eqmot1}) that for the free particle $v$ and $\bar v$ are simply constant in time, and are directly related to the momentum. (The former observation follows from the first two lines of (\ref{eqmot1}), whereas the latter follows from the third and fourth lines). This again confirms the picture found above, namely that for the free particle the deformation in $v$ depends linearly on the momentum, which was also the conclusion reached in \cite{susskind}. In Section \ref{freepart} we will investigate the quantum mechanical free particle, also in lieu of the connection between momentum and extent. \\

Clearly the arguments above support the notion that, on the classical level, $v$ may be viewed as describing the extent of a composite. This was demonstrated explicitly through introduction of the the two coupled charges in a magnetic field, and subsequent analysis of corrections to the standard equations of motion and conserved energy. We shall proceed by taking this view as a point of departure for the physical interpretation of our non-commutative quantum system. It will be demonstrated that said view is indeed also a natural one on the quantum level.

\section{Constraints and differential operators on $\overlapr{z,v}{\psi}$}

Due to its corresponding local probability description, the basis $\ketr{z,v}$ is mathematically convenient for the representation of states $\ketr{\psi}$ of a non-commutative quantum systems in terms of overlaps $\overlapr{z,v}{\psi}$. From (\ref{zv2}) we see that the bra in $\mathcal{H}_{q}^{*}$ dual to $\ketr{z,v}$ is

\beqa
\brar{z,v} &=& \ket{z+v} \bra{z} e^{\frac{1}{2}(\bar{v}z-\bar{z}v)}  \nonumber \\
&=& e^{-[z\bar{z} + \bar{z}v + \frac{1}{2}v\bar{v}]} e^{(z+v)\bd}\ket{0}\bra{0} e^{\bar{z}b}.
\label{zvdual}
\eeqa
Since all operators on $\hq$ may be written as functions of the bosonic operators (\ref{BBD}), we now proceed to show that the action of these operators on a state $\ketr{\psi}$ may be described in terms of differential operators acting on the overlap $\overlapr{z,v}{\psi}$. Using the notation for left- and right acting operators set out in (\ref{OLR}), we have
\beqa
\matrelr{z,v}{\bld}{\psi} & = & e^{\frac{1}{2}(\bar{v}z-\bar{z}v)} \matrel{z}{\bd\psi}{z+v} = \bar{z}\overlapr{z,v}{\psi}, \nonumber \\
\matrelr{z,v}{\bl}{\psi}  & = & e^{\frac{1}{2}(\bar{v}z-\bar{z}v)} \matrel{z}{b\psi}{z+v} \,\,= (\diffpa{\bar{z}}+z+v)\overlapr{z,v}{\psi}, \nonumber \\
\matrelr{z,v}{\br}{\psi}  & = & e^{\frac{1}{2}(\bar{v}z-\bar{z}v)} \matrel{z}{\psi b}{z+v} \,\,= (z+v)\overlapr{z,v}{\psi}, \quad\quad\quad\textnormal{and}\nonumber \\
\matrelr{z,v}{\brd}{\psi} & = & e^{\frac{1}{2}(\bar{v}z-\bar{z}v)} \matrel{z}{\psi\bd}{z+v} = (\diffpa{v}+\bar{z}+\frac{\bar{v}}{2})\overlapr{z,v}{\psi}\nl &=& (\diffpa{z}+\bar{z}) \overlapr{z,v}{\psi}.
\label{ladders}
\eeqa
Next we note from equation (\ref{zvdual}) that the functions $\overlapr{z,v}{\psi}$ must obey the following set of constraints,
\beqa
&&\left(\diffpa{\bar{v}}+\frac{v}{2}\right )\overlapr{z,v}{\psi}= 0,\label{constr1}\\
&&\left(\diffpa{z}-\diffpa{v}- \frac{\bar{v}}{2}\right) \overlapr{z,v}{\psi}=0.
\label{constr2}
\eeqa
Since these constraints are a consequence of the choice of basis, they must hold for \emph{all} states $\psi$. Consequently there is a restriction on which functions $\overlapr{z,v}{\psi}$ are physical in this basis. We shall refer to this subspace of functions as the physical subspace. As stated earlier, our intention is to represent all operators that are functions of bosonic operators in terms of differential operators. In this light, it should be noted not only that the constraints (\ref{constr1}) and (\ref{constr2}) commute with each other, but also that the differential operators associated with the creation and annihilation operators in (\ref{ladders}) all commute with the constraints. Consequently the physical subspace of functions is left invariant under the action of aforementioned differential operators, and thus also under the action of the differential operator representation in this basis of any operator that is a function of those in (\ref{ladders}). This implies that we may implement the constraints strongly on the physical subspace, and we shall do so in subsequent analyses. A further consequence of the constraints is that the differential operator representation of a particular operator is not unique on the physical subspace, since we may employ (\ref{constr1}) and (\ref{constr2}) to rewrite such a representation. Indeed, we shall make use of this feature repeatedly in sections to follow. It should be noted, however, that this procedure is only valid on the physical subspace, and must be implemented with great care.

As stated, (\ref{ladders}) provides us with a useful ``dictionary'' to represent any operator that is a function of the bosonic operators in terms of derivatives acting on $\overlapr{z,v}{\psi}$. In the subsequent sections we analyse the angular momentum operator and the Hamiltonians of the free particle and the generalised harmonic oscillator by looking at their representations and eigenstates in the basis (\ref{zv2}).

\section{Angular momentum}
\label{ang}

In \cite{jopa} it was shown that the generator of rotations in this non-commutative quantum mechanical formalism is
\beq
L =  X_L P_y - Y_L P_x + \frac{\theta}{2\hbar} PP^\ddagger.
\label{Lz}
\eeq
The first two terms are the familiar $\vec r \times \vec p$ part of the angular momentum operator. The second term arises due to the non-commutativity of co-ordinates and also ensures that angular momentum is a conserved quantity for the free particle, as required.
Since our ``recipe book'' (\ref{ladders}) allows us to represent functions of the creation- and annihilation operators, the next step is to rewrite (\ref{Lz}) in terms of these operators. To do so, we note from (\ref{schnc}) that
\beq
X_L=\sqrt{2\theta}(\bl+\bld), \quad Y_L=-i \sqrt{2\theta}(\bl-\bld), \quad P_x=\frac{\hbar}{\theta}(Y_L-Y_R), \quad \textnormal{and} \quad P_y=\frac{\hbar}{\theta}(X_R-X_L).
\label{leftrightmomenta}
\eeq
Inserting this into (\ref{Lz}) we find that
\beq
L = \hbar\left(\br\brd-\bld\bl\right),
\label{Lzladder}
\eeq
where $\br\brd=(B^\ddagger B)_R$ is the right number operator. (The order of operators is indeed correct here, since for any two operators $A$ and $B$ we have that $[AB]_R\psi=\psi A B = B_R A_R \psi$). We may now write the action of this angular momentum operator in the basis (\ref{zv2}) as a differential operator (denoted as $\hat{L}$) by making use of the relevant associations from (\ref{ladders}):
\beqa
\hat{L} &=& \hbar \left[(z+v)\left(\diffpa{v}+\bar{z}+\frac{\bar{v}}{2}\right)-\bar{z}\left(\diffpa{\bar{z}}+z+v\right)\right] \nonumber \\
&=& \hbar \left[ v\diffpa{v}+z\diffpa{v}+\frac{|v|^2}{2}+\frac{\bar{v}z}{2}-\bar{z}\diffpa{\bar{z}}\right].
\label{Lzdiff}
\eeqa
Although this representation is unique on the full space, it can be cast in different forms on the physical subspace using the constraints (\ref{constr1}), (\ref{constr2}). To illustrate this, we note that (\ref{Lzdiff}) may be rewritten as
\beqa
\hat{L} &=& \hbar \left[z\diffpa{z}-\bar{z}\diffpa{\bar{z}}+v\diffpa{v}-\bar{v}\diffpa{\bar{v}} \right]. \nonumber \\
&=& \hat{L}_z + \hat{L}_v
\label{LzLv}
\eeqa
on the physical subspace. This particular form of $\hat L$ is manifestly Hermitian but, as stated, it may only be applied to elements of the physical subspace and is not valid on the unconstrained function space. Furthermore, if we view $\hat{L}_z=\hbar\left(z\diffpa{z}-\bar{z}\diffpa{\bar{z}}\right)$ and $\hat{L}_v=\hbar\left(v\diffpa{v}-\bar{v}\diffpa{\bar{v}}\right)$ as an orbital angular momentum and an intrinsic angular momentum respectively, we note that there are two contributions to the total angular momentum. This point of view is not unreasonable if we interpret $z$ and $v$ as position and local spatial variations of the state, respectively. Taking this view as a point of departure, we thus see the explicit split of total angular momentum into orbital and intrinsic angular momentum. This interpretation is clearly in line with the notion of an extended or structured object. One should, however, be cautious in applying this interpretation, since (\ref{LzLv}) acts only on the constrained physical subspace. It would be wrong to think that $\hat{L}_z$ and $\hat{L}_v$ are independent operators, i.e. that one could define states on the physical subspace that are simultaneous eigenstates of $\hat{L}_z$ and $\hat{L}_v$. In fact, in lieu of the constraints (\ref{constr1}), (\ref{constr2}) it becomes clear that these two operators do not commute on the physical subspace. Consequently it is not surprising that such physical simultaneous eigenstates of $\hat{L}_z$ and $\hat{L}_v$ do not exist. To shed some light on this matter, let us consider eigenstates of the total angular momentum operator.

\begin{figure}[tb]
\begin{center}
\leavevmode
\includegraphics[width=0.6\textwidth]{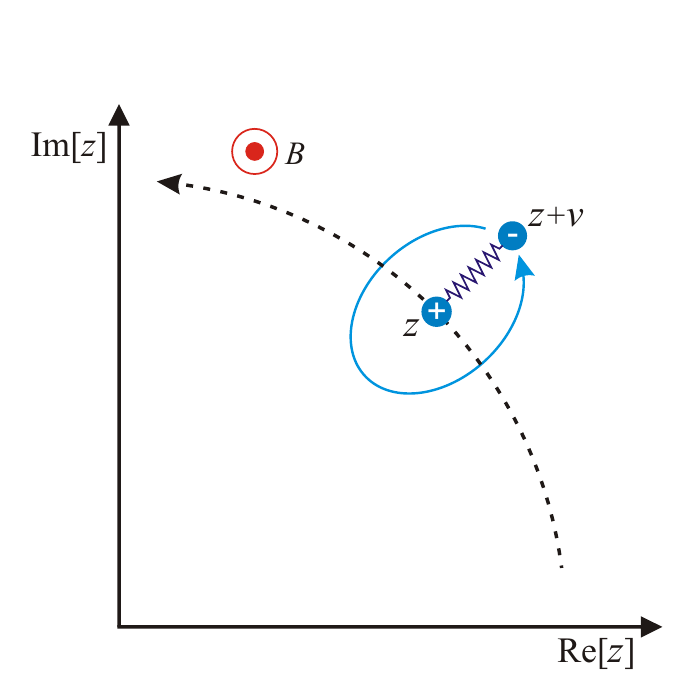}
\end{center}
\caption{A schematic showing the two-charge (harmonically coupled) composite, whose orbital motion affects its shape deformation and vice versa.}
\label{fig:chargesimage}
\end{figure}

From the form (\ref{Lzladder}) of the angular momentum, it is clear that the states
\beq
\ketr{l} \equiv \sum_{n=0}^{\infty} \alpha_n \ket{n}\bra{n+l}
\label{Leigenstates}
\eeq
are eigenstates of $L$, since
\beqa
L\ketr{l} &=& \hbar\left(\br\brd-\bld\bl\right)\ketr{l} \nonumber \\
&=& \hbar\sum_{n=0}^{\infty} \alpha_n \left( \ket{n}\bra{n+l}\bd b - \bd b\ket{n}\bra{n+l} \right) \nonumber \\
&=& \hbar l \ketr{l}.
\eeqa
To represent such a state in the basis (\ref{zv2}) we require its overlap with the bra (\ref{zvdual}), namely
\beqa
\overlapr{z,v}{l} &=& e^{\frac{1}{2}(\bar{v}z-\bar{z}v)} \sum_{n=0}^{\infty} \alpha_n \tr{[\ket{z+v}\bra{z}]\ket{n}\bra{n+l}} \nonumber \\
&=& e^{\frac{1}{2}(\bar{v}z-\bar{z}v)} \sum_{n=0}^{\infty} \alpha_n \overlap{z}{n}\overlap{n+l}{z+v} \nonumber \\
&=& e^{\frac{1}{2}(\bar{v}z-\bar{z}v)} e^{-(|z|^2+|z+v|^2)/2}\sum_{n=0}^{\infty} \alpha_n \frac{\bar{z}^n(z+v)^n}{n!}\frac{(z+v)^l}{\sqrt{(n+l)!/n!}}.
\label{zvl}
\eeqa
By construction this overlap is an element of the physical subspace, and thus it is an eigenfunction of both total angular momentum differential operators (\ref{Lzdiff}) and (\ref{LzLv}). Note, however, that the variables $z$ and $v$ do not decouple in this ``wave function''. We see here explicitly that although (\ref{zvl}) is an eigenstate of total angular momentum, it is not a simultaneous eigenstate of $\hat{L}_z$ and $\hat{L}_v$. As stated, due to the constraints it is impossible to find such a state on the physical subspace of $\mathcal{H}_q$ --- the requirement of physicality prevents the decoupling of $z$ and $v$ as is seen in (\ref{zvl}). A physical consequence hereof is that the quantities that may be interpreted as orbital and intrinsic angular momentum, respectively, are not independent. If we consider the clear connection between momentum (motion of the classical composite) and shape deformation seen in Section \ref{class}, this result is reasonable also from a physical point of view. The implication is simply that orbital motion affects shape deformation and consequently intrinsic angular momentum, and vice versa. We infer that $z$ and $v$ cannot be interpreted as degrees of freedom of a rigid body. This too is in line with the results of \cite{susskind}. For a schematic of this scenario, see Figure \ref{fig:chargesimage}
\vspace{0.cm}

\section{Free particle}
\label{freepart}
\def\aa {\sqrt{\frac{\theta}{2\pi\hbar^{2}}}}
\def\ab {\frac{\theta}{4\hbar^{2}}}
\def\ac {\frac{i}{\hbar} \sqrt{\frac{\theta}{2}}}
The Hamiltonian of the free particle is simply
\beq
H_{free} = \frac{P^\ddagger P}{2m}= -\frac{\hbar^2}{m\theta}[\bld-\brd]\left[\br-\bl\right],
\label{Hfree}
\eeq
where we have used definition (\ref{PPD}) of the complex momenta. Again we write the action of (\ref{Hfree}) on a state $\ketr \psi$ as a differential operator in the basis (\ref{zv2}) according to (\ref{ladders}),
\beqa
\matrelr{z,v}{H_{free}}{\psi} &=& -\frac{\hbar^2}{m\theta} \left[ \bar{z}- (\diffpa{z}+\bar{z})\right] \left[(z+v)-(\diffpa{\bar{z}}+z+v)\right]
\overlapr{z,v}{\psi} \nonumber \\
&=& -\frac{\hbar^2}{m\theta} \frac{\partial^2}{\partial z\partial \bar{z}}\overlapr{z,v}{\psi}\equiv \hat{H}_{free}\overlapr{z,v}{\psi}.
\label{freeHdiff}
\eeqa
Note that the operator $\hat{H}_{free}$ is independent of $v$, which implies a complete decoupling between the positional degrees of freedom ($z$) and those that pertain to additional structure ($v$) for the free particle. This is not surprising since we know that non-commutativity has no effect on a free particle, as was found in \cite{jopa}.

Next we consider the eigenstates of momentum as given in \cite{jopa},
\beq
\ketr{\psi_{k}} =\aa e^{\ac(\bar{k}b+k\bd)}= \aa e^{-\ab |k|^{2}}e^{\ac k\bd} e^{\ac \bar{k}b}.
\label{psik}
\eeq
The normalisation prefactor is chosen thus so that these states form a complete set of basis states on $\hq$. This will be proven and used at a later stage. These states are analogues of plane waves from standard quantum mechanics (as is seen from definition (\ref{bbd})), and are clearly eigenstates of the complex momenta (\ref{PPD}),
\beqa
P \ketr{\psi_{k}}&=& i\hbar \sqrt{\frac{2}{\theta}}\aa e^{-\ab |k|^{2}}\left[e^{\ac k\bd} e^{\ac \bar{k}b},\;b\,\right]=i\hbar \sqrt{\frac{2}{\theta}}k\ketr{\psi_k},\nl
P^\ddagger\ketr{\psi_{k}} &=& i\hbar\sqrt{\frac{2}{\theta}}\aa e^{-\ab |k|^{2}}\left[b^\dagger,e^{\ac k\bd} e^{\ac \bar{k}b}\right]=i\hbar \sqrt{\frac{2}{\theta}}\bar k \ketr{\psi_k},
\eeqa
and consequently also eigenstates of the free particle Hamiltonian (\ref{Hfree}). The overlap of such a momentum state with a basis element (\ref{zv2}) is
\beqa
\overlapr{z,v}{\psi_{k}} &=& \aa e^{-\ab |k|^{2}}e^{\frac{1}{2}(\bar{v}z-\bar{z}v)}\bra{z} e^{\ac k\bd} e^{\ac \bar{k}b}\ket{z+v}\nonumber\\
&=& \aa e^{-\ab |k|^{2}} e^{\ac [k\bar{z}+\bar{k}(z+v)]} e^{-\frac{1}{2}|v|^2}.
\eeqa
These overlaps are, by construction, eigenstates of the differential operator representation (\ref{freeHdiff}) of the free particle Hamiltonian (\ref{Hfree}). As expected, the $z$ and $v$ degrees of freedom decouple in this wave function. We can find the probability distribution in $z$ and $v$ for the state $\ketr{\psi_{k}}$ using the POVM (\ref{pizv}),
\beqa
P(z,v) &=& \matrelr{\psi_{k}}{\pi_{z,v}}{\psi_{k}}\nonumber \\
&=& \overlapr{\psi_{k}}{z,v} \overlapr{z,v}{\psi_{k}} \nonumber \\
&=& \frac{\theta}{2\pi\hbar^{2}} e^{-\frac{\theta}{2\hbar^{2}}|k|^{2}}e^{\ac [\bar{k}v - \bar{v}k]} e^{-\modsq{v}} \nl
&=& \frac{\theta}{2\pi\hbar^{2}} e^{-|\ac k - v|^2}.
\label{Pzvfree}
\eeqa
The first evident feature of distribution (\ref{Pzvfree}) is that all dependence on $z$ has disappeared. This is, of course, to be expected and simply implies that the dynamics of the average position, or ``guiding center'' is that of a free particle, as confirmed by (\ref{freeHdiff}). A measurement of position, which does not enquire about any other possible structure, will therefore yield equal probabilities everywhere.
The Gaussian $k$-dependence implies a regularization of high momenta. This is simply a consequence of the existence of a short length scale $\sqrt\theta$, since a minimal length scale implies (through a Fourier transformation) that high momenta are restricted --- a result that was also found for the non-commutative spherical well in \cite{ncwell}. Next we note that, for a fixed value of $k,\bar{k}$, the term $e^{\ac [\bar{k}v - \bar{v}k]}$ represents a momentum-dependent stretching of the distribution in $v$. This stretching is perpendicular to the direction of motion, which has the physical implication that a measurement of the distribution around the center $z$ through the implementation of the POVM (\ref{pizv}) will yield an asymmetrical momentum-dependent distribution, very much as was found in \cite{susskind}. The Gaussian $v$-dependence shows that the spatial distribution around $z$ is confined on the length scale set by $\theta$.  Note that this is a generic feature, which does not depend on dynamics as the Gaussian factor in the wave-function is a consequence of the constraint (\ref{constr1}). We thus expect the distribution in $v$ always to be confined to a length scale set by $\theta$, regardless of the particular dynamics, while said dynamics will set the length scale associated with the average position $z$.  We shall indeed see this explicitly for the harmonic oscillator discussed below. In the case of the free particle there is of course no length associated with the average position $z$. One could also view the Gaussian dependence of the wave function on $v$ as arising from harmonic dynamics for $v$ with oscillator length $\sqrt{2\theta}$, which, for small values of $\theta$, corresponds to a very stiff spring constants. This is, of course, precisely the picture that emerged from the corresponding classical theory discussed in Section \ref{class}.

\section{Harmonic oscillator}
\label{harmos}

The harmonic oscillator Hamiltonian discussed in \cite{jopa} was

\beq
H = \frac{1}{2m} PP^\ddagger + \frac{1}{2}m\omega^2 ({X}_L^2 +{Y}_L^2),
\label{Hharm1}
\eeq
where we note that the harmonic interaction may be rewritten in terms of the creation- and annihilation operators (\ref{BBD}) through $\bld \bl = \frac{1}{2\theta}({X}_L^2 +{Y}_L^2)+\frac 1 2$. Note that we shall omit the factor of $\frac 1 2$ in the Hamiltonian below, since it constitutes a constant contribution to the energy. We proceed by generalising this Hamiltonian slightly through the addition of a similar harmonic term with right action, which yields the Hamiltonian that we shall consider for the rest of this analysis:
\beqa
H_{h.o.} &=& \frac{1}{2m} PP^\ddagger + m\theta\omega_L^2 (\bld \bl) + m\theta \omega_R^2 (\br \brd) \nonumber \\
&=& \alpha \bld \bl + \beta \brd \br - \gamma (\bld \br + \brd \bl) - m\theta \omega_R^2,
\label{Hharm2}
\eeqa
where we identify
\beq
\alpha = \frac{\hbar^2}{m\theta}+m\theta\omega_L^2, \quad
\beta  = \frac{\hbar^2}{m\theta}+m\theta \omega_R^2, \quad
\gamma = \frac{\hbar^2}{m\theta}.
\label{abg}
\eeq
Returning to (\ref{leftrightmomenta}), we note that the right action term may also be rewritten in terms of momenta and left co-ordinates. Consequently one may also interpret the generalised Hamiltonian (\ref{Hharm2}) as a gauged harmonic oscillator Hamiltonian with an added magnetic field. Next we wish to diagonalise this Hamiltonian, i.e., find its eigenstates. The addition of the right action term implies that we cannot follow the diagonalisation procedure discussed in \cite{jopa}. We digress briefly to describe the diagonalisation of (\ref{Hharm2}) through the construction of a Bogoliubov transformation. This transformation introduces new ladder operators of the form
\beq
\left(
  \begin{array}{c}
    A_1 \\
    A_1^\ddagger \\
    A_2 \\
    A_2^\ddagger \\
  \end{array}
\right)
=
\left(
  \begin{array}{cccc}
    \func{cosh}{\phi} & 0 & \func{sinh}{\phi} & 0 \\
    0 & \func{cosh}{\phi} & 0 & \func{sinh}{\phi} \\
    \func{sinh}{\phi} & 0 & \func{cosh}{\phi} & 0 \\
    0 & \func{sinh}{\phi} & 0 & \func{cosh}{\phi} \\
  \end{array}
\right)
\left(
  \begin{array}{c}
    \bl \\
    \bld \\
    \br \\
    \brd \\
  \end{array}
\right),
\label{bogol1}
\eeq
and preserves the commutation relations of $B_L$, $B_L^\ddagger$, $B_R$ and $B_R^\ddagger$, i.e.,
\beqa
&[\bl,\bld]=1;\,[\br,\brd]=-1;\,[\bl,\br]=[\bl,\brd]=0& \nl
&\Downarrow&\nl
&[A_1,A_1^\ddagger]=1;\,[A_2,A_2^\ddagger]=-1; [A_1,A_2]=[A_1,A_2^\ddagger]=0.&
\eeqa
Insisting on a diagonal form of (\ref{Hharm2}) in terms of these new operators fixes the rotation parameter $\phi$ on
\def\rootA{\sqrt {\frac{(\lambda_L+\lambda_R)(4 \hbar^2+ m\theta[\lambda_L+\lambda_R])}{m\theta}}}
\def\rootB{\sqrt {(\omega_L^2+\omega_R^2)[4\hbar^2+m^2\theta^2(\omega_L^2+\omega_R^2)]}}
\def\gamA{1+\frac{m\theta}{2\hbar^2}\left[ \lambda_L+\lambda_R -\rootA  \right]}
\def\gamB{1+\frac{m\theta}{2\hbar^2}\left[ m\theta(\omega_L^2+\omega_R^2) -\rootB  \right]}
\beq
\phi = -\func{arctanh}{\Gamma},
\label{phi}
\eeq
with
\beq
\Gamma=\gamB
\eeq
Under (\ref{phi}), the inversion of (\ref{bogol1}) and subsequent substitution into (\ref{Hharm2}) yield the diagonalised Hamiltonian
\beq
H_{h.o.} = K_1 A_1^\ddagger A_1  + K_2 A_2 A_2^\ddagger + (K_2 - m\theta\omega_R^2)\;
\label{Hharm3}
\eeq
where we denoted
\beqa
K_1 &=& \frac{1}{2} \left[ m\theta\omega_L^2-m\theta\omega_R^2+ \rootB \right], \nonumber \\
K_2 &=& \frac{1}{2} \left[ m\theta\omega_R^2-m\theta\omega_L^2+ \rootB \right].
\eeqa
It is now a simple matter to see that the spectrum of (\ref{Hharm2}) is
\beq
E_{n_1,n_2} = n_1 K_1 + (n_2+1) K_2 - m\theta\omega_R^2.
\label{spectrum}
\eeq
Next we construct the vacuum solution, $\ketr{0}$. It is required that this state is annihilated by the relevant operators, namely
\beqa
A_1\ketr{0} = 0 &\Rightarrow& [ \cosh{(\phi)} \bl + \sinh{(\phi)}\br ] \ketr{0} = 0, \quad \textnormal{and} \nonumber \\
A_2^\ddagger\ketr{0} = 0 &\Rightarrow& [ \sinh{(\phi)} \bld + \cosh{(\phi)}\brd ] \ketr{0} = 0.
\label{annihilate}
\eeqa
Since we know that $\bl\ket{0}\bra{0}=\brd\ket{0}\bra{0}=0$, let us postulate that $\ketr{0}=\mathcal{N}e^{\xi\bld\br}\ket{0}\bra{0}$, in which case we find
\beqa
A_1\ketr{0} &=&\mathcal{N}  \left( \cosh{(\phi)}\{[\bl,e^{\xi\bld\br}]+e^{\xi\bld\br}\bl\} +\sinh{(\phi)}\br e^{\xi\bld\br}\right)\ket{0}\bra{0} \nonumber\\
&=& \left( \cosh{(\phi)}\xi\br + \sinh{(\phi)}\br \right)\ketr{0} \quad \textnormal{and}\nonumber \\
A_2^\ddagger\ketr{0} &=&\mathcal{N} \left( \sinh{(\phi)}\bld e^{\xi\bld\br} +\cosh{(\phi)}\{[\brd,e^{\xi\bld\br}]+e^{\xi\bld\br}\brd\} \right) \ket{0}\bra{0} \nonumber \\
&=& \left( \sinh{(\phi)}\bld + \cosh{(\phi)}\xi\bld \right)\ketr{0}
\eeqa
Clearly (\ref{annihilate}) is satisfied if we choose $\xi = -\frac{\sinh{\phi}}{\cosh{\phi}}=-\tanh{\phi}$, i.e. when $\ketr{0}=\mathcal{N}e^{\Gamma\bld\br}\ket{0}\bra{0}$ (see (\ref{phi})). For the normalisation of the ground state we note that
\beqa
\overlapr{0}{0} &=& \mathcal{N}^2\, \tr{\left[e^{\Gamma\bld\br}\ket{0}\bra{0}\right]^\ddagger e^{\Gamma\bld\br}\ket{0}\bra{0}} \nonumber \\
&=& \mathcal{N}^2\sum_{n=0}^\infty \sum_{m=0}^\infty \Gamma^{n+m} \tr{\ket{n}\overlap{n}{m}\bra{m}} \nonumber \\
&=& \mathcal{N}^2\sum_{n=0}^\infty \Gamma^{2n} \nonumber \\
&=& \frac{\mathcal{N}^2}{1-\Gamma^2},
\eeqa
where the condition $|\Gamma|<1$ is automatically satisfied due to (\ref{phi}). Thus the correctly normalized ground state is
\beq
\ketr{\psi_0}=\sqrt{1-\Gamma^2}e^{\Gamma\bld\br}\ket{0}\bra{0}.
\label{ground}
\eeq
Finally, excited states can be constructed by applying the appropriate ladder operators from (\ref{bogol1}):
\beq
\ketr{n_1,n_2}_{h.o.}=(A_1^\ddagger)^{n_1}(A_2)^{n_2}\ketr{\psi_0}.
\eeq
In the limit $\omega_R\rightarrow 0$ the above results reduce to those of \cite{jopa}. Consider, for instance, the probability distribution in position for the ground state (\ref{ground}). To do this we note that for $\ketr{z}=\ket{z}\bra{z}$,
\beqa
\overlapr{z}{\psi_0} &=& \sqrt{1-\Gamma^2}\sum_{n=0}^\infty\Gamma^n e^{-|z|^2}\frac{|z|^{2n}}{n!} \nonumber\\
&=& \sqrt{1-\Gamma^2}e^{|z|^2(\Gamma-1)}.
\eeqa
Thus
\beqa
P(z) &=& \overlapr{\psi_0}{z} \star \overlapr{z}{\psi_0} \nonumber\\
&=& (1-\Gamma^2)\sum_{n=0}^\infty\frac{(z\bar{z})^n(\Gamma-1)^{2n}}{n!}e^{2|z|^2(\Gamma-1)} \nonumber\\
&=& (1-\Gamma^2)e^{-(1-\Gamma^2)|z|^2}.
\eeqa
In the limit $\omega_R\rightarrow 0$ this agrees with the distribution found in \cite{jopa}.

Next we use (\ref{ladders}) to find the representation of the harmonic oscillator Hamiltonian (\ref{Hharm2}) in the basis (\ref{zv2}):
\beqa
\matrelr{z,v}{H_{h.o.}}{\psi} &=& \matrelr{z,v}{\frac{1}{2m} PP^\ddagger + m\theta\omega_L^2 (\bld \bl) + m\theta \omega_R^2 (\br \brd)}{\psi} \nonumber \\
&=& \left[-\frac{\hbar^2}{m\theta} \frac{\partial^2}{\partial z\partial {\bar{z}}}+m\theta\omega_L^2\bar{z}(\diffpa{\bar{z}}+z+v)
+m\theta\omega_R^2(z+v)(\diffpa{v}+\bar{z}+\frac{\bar{v}}{2})\right] \overlapr{z,v}{\psi} \nonumber \\
&:=& \hat{H}_{h.o.} \overlapr{z,v}{\psi}.
\label{Hharmdiff}
\eeqa
We note that the operator $\hat{H}_{h.o.}$ is only Hermitian on the physical function space restricted by the constraints (\ref{constr1}) and (\ref{constr2}), and that its particular form in (\ref{Hharmdiff}) is again not unique on this space due to the constraints. Furthermore, it is easy to check that the total angular momentum operator (\ref{LzLv}) commutes with the above Hamiltonian as desired.

As mentioned above, the constraints (\ref{constr1}), (\ref{constr2}) allow the rewriting of (\ref{Hharmdiff}) in many equivalent forms on the physical subspace. One particular form, namely the manifestly Hermitian form, reflects the physics more explicitly. Through an appropriate use of constraints the Hamiltonian (\ref{Hharmdiff}) can indeed be rewritten as
\beqa
&&\hat{H}_{h.o.}= -\frac{\hbar^2}{m\theta} \frac{\partial^2}{\partial z\partial {\bar{z}}}+m\theta\omega_L^2\left[|z|^2+\bar{z}(\diffpa{\bar{z}}+\frac{v}{2}-\diffpa{\bar v})+z(-\diffpa{z}+\frac{\bar v}{2}+\diffpa{v})\right]+\nonumber\\
&&m\theta\omega_R^2\left[(z+\frac{v}{2}-\diffpa{\bar v})(\bar{z}+\diffpa{v}+\frac{\bar{v}}{2})\right]\nonumber\\
&&=-\frac{\hbar^2}{m\theta}\frac{\partial^2}{\partial z\partial {\bar{z}}}+m\theta(\omega_L^2+\omega_R^2)|z|^2-m\theta\omega_L^2\left(z\diffpa{z}-\bar{z}\diffpa{\bar z}\right) +m\theta\omega_R^2\left[(\diffpa{ v}+\frac{\bar v}{2})(-\diffpa{\bar v}+\frac{v}{2})-1\right]\nonumber\\
&&+m\theta(\omega_L^2+\omega_R^2)\left(\bar z(-\diffpa{\bar v}+\frac{v}{2})+z(\diffpa{v}+\frac{\bar v}{2})\right).\nonumber\\
\label{Hharmdiff1}
\eeqa
The different contributions in this Hamiltonian have clear physical meanings. The first two terms represent a normal harmonic oscillator whose frequency is shifted by the right frequency, i.e., if we impose $\omega_R\rightarrow0$ this is simply the standard harmonic oscillator Hamiltonian. The third term reflects the standard type of ``Zeeman term'' which has a clear angular momentum dependence. It is this term that induces the well-known time reversal symmetry breaking, since ``forward'' and ``backward'' angular momentum states (of which one is obtained by time-reversing the other) do not have equal energy. The fourth term represents a ``Landau'' Hamiltonian for the variable $v$ with energy scale set by $\omega_R$. This again supports the notions that the right sector displays harmonic dynamics and that the right action term in the Hamiltonian can also be rewritten in terms of a standard gauged magnetic field term, i.e., in terms of left acting co-ordinates and momenta. The last term represents the expected coupling between the variable $v$, describing the local spatial distribution of the state, and the average position $z$, implying that the local spatial distribution will be position-dependent. Again we note that due to the constraint (\ref{constr1}) the wave function $\left(z,v|\psi\right)$ must always contain a Gaussian $e^{-\frac{|v|^2}{2}}$, which implies that this dimensionless parameter is of order $v\sim 1$, i.e., the dimensionful variable $v\sim \sqrt{\theta}$.  Introducing the dimensionful variable $z^\prime=\sqrt{2\theta}z$ one immediately sees from this that the third through last terms are all higher order in $\theta$ and will vanish in the commutative limit to yield the standard commutative harmonic oscillator. To recap, we observe that the form (\ref{Hharmdiff}) of the Hamiltonian which is unique on the entire space, is not Hermitian. Since the physical states are those that are annihilated by the constraints, we could thus find the solutions to this non-Hermitian Hamiltonian on the whole space, and select only the physical ones (since the constraints and the Hamiltonian commute). This simply amounts to selecting the ``lowest Landau level'' states of the right sector. The interaction term could then, for instance, be solved perturbatively.

Lastly, let us look at the representation of the ground state (\ref{ground}) in the basis (\ref{zv2}):
\beqa
\overlapr{z,v}{\psi_0} &=& \sqrt{1-\Gamma^2}e^{\frac{1}{2}(\bar{z}v-\bar{v}z)}\tr{\ket{z+v}\bra{z} \left[e^{\Gamma\bld\br}\ket{0}\bra{0}\right]} \nonumber\\
&=& \sqrt{1-\Gamma^2}e^{\frac{1}{2}(\bar{z}v-\bar{v}z)} \sum_n^\infty \Gamma^n \overlap{z}{n}\overlap{n}{z+v} \nonumber \\
&=& \sqrt{1-\Gamma^2}e^{\frac{1}{2}(\bar{z}v-\bar{v}z)} e^{\Gamma \bar{z}(z+v)} e^{-\frac{1}{2}(|z|^2+|z+v|^2)}
\eeqa
This implies that the probability distribution in $z$ and $v$ for the ground state is
\beqa
\label{Pzvharm}
P(z,v) &=& \matrelr{\psi_0}{\pi_{z,v}}{\psi_0}\nonumber \\
&=& \overlapr{\psi_{0}}{z,v} \overlapr{z,v}{\psi_{0}} \nonumber \\
&=& (1-\Gamma^2) e^{\Gamma(2|z|^2+\bar{z}v+\bar{v}z))} e^{-(2|z|^2+|v|^2+\bar{z}v+\bar{v}z)} \nonumber \\
&=& (1-\Gamma^2)\underbrace{e^{-|v|^2}} \underbrace{e^{-2(1-\Gamma)|z|^2}} \underbrace{e^{-(1-\Gamma)(\bar{z}v+\bar{v}z)}} \\
&{}& \qquad\qquad\,\,\textnormal{(i)}\, \qquad\,\;\textnormal{(ii)} \qquad\quad \;\;\;\, \textnormal{(iii)} \nonumber
\eeqa
Let us first investigate this distribution in the standard non-commutative harmonic oscillator limit, i.e. where $\omega_R=0$. For this purpose we define two length scales,
\beqa
\ell_\theta &=& \sqrt{2\theta} \nonumber \\
\ell_{\omega_L} &=& \sqrt{\frac{2\hbar}{m\omega_L}},
\eeqa
where $\ell_{\omega_L}$ is just the standard harmonic oscillator length scale. Noting that both $z$ and $v$ are dimensionless variables, i.e. $z=\frac{1}{\sqrt{2\theta}}(x+iy)$ (and similarly for $v$), we see that the Gaussian (i) in (\ref{Pzvharm}) decays on a length scale of $\ell_\theta$. If we associate $v$ with spatial extent, this is to be expected as the scale for the local spatial extent must be set by the non-commutative parameter. As already remarked this behaviour is quite generic and a consequence of the constraints on the wave function rather than the dynamics. This is also important to ensure that the variable $v$ couples weakly to the variable $z$ for small $\theta$ and decouples in the commutative limit.

From (\ref{phi}) it is clear that
\beq
\Gamma |_{\omega_R=0} = 1+\frac{m\theta}{2\hbar^2}\left[ m\theta\omega_L^2 -\sqrt{\omega_L^2[4\hbar^2+m^2\theta^2\omega_L^2]}  \right],
\eeq
and thus
\beq
1-\Gamma |_{\omega_R=0} = -\left(\frac{\ell_\theta}{\ell_{\omega_L}}\right)^2\left[2 \left(\frac{\ell_\theta}{\ell_{\omega_L}}\right)^2 - 2\sqrt{1+\left(\frac{\ell_\theta}{\ell_{\omega_L}}\right)^4}\right].
\eeq
Since $v = \frac{1}{\ell_\theta}(v_x+iv_y)$, it is clear that under these assumptions the Gaussian (ii) in (\ref{Pzvharm}) decays on a length scale of $\ell_{\omega_L}$, with a further dependence on the ratio $\left(\frac{\ell_\theta}{\ell_{\omega_L}}\right)$. Comparing this to the case of the commutative harmonic oscillator, where the ground state wave function decays on a length scale of $\ell_\omega=\sqrt{\frac{2\hbar}{m\omega}}$, this also makes sense: the variable $z$ is associated with the position of a particle moving in a harmonic potential with strength set by $\omega_L$. Finally, we observe that term (iii) in (\ref{Pzvharm}) represents the expected position dependent deformation of the distribution $P(z,v)$.  Note that when $\omega_L=\omega_R=0$ this term vanishes and, as was found for the free particle, there is a decoupling.

In conclusion, the representation of the ground state (\ref{ground}) for the case $\omega_R=0$ in the basis (\ref{zv2}) shows explicitly that there are two length scales involved in the problem: the fundamental harmonic oscillator length scale \textit{as well as} the length scale set by the non-commutative parameter $\theta$. As discussed in Section \ref{class}, it is of course generic that the particular dynamics set the positional length scale of a problem. In the case where $\omega_R\neq0$, the decay of the Gaussian term (iii) in (\ref{Pzvharm}) would be governed by two length scales: $\ell_{\omega_L}$ and $\ell_{\omega_R}$.


\chapter{THE BASIS $\ketr{z,n}\equiv T(z)\ket 0 \bra n $}
\label{znchapter}

\section{A basis with a discrete right sector label  --- interpretation and probability distribution}
\label{znsection1}
Of course the particular degree of freedom introduced to decompose the star product need not be a continuous state label. We could just as
well choose a discrete basis to label the right sector and still have a consistent local description for position measurements. To do this,
let us revisit the form (\ref{Lzladder}) of the total angular momentum operator and its most general eigenstates (\ref{Leigenstates}). It
is clear that the state
\beq
\ketr{m,n} \equiv \ket m \bra n ,
\label{mn}
\eeq
with $\ket n \in \hc$ as in (\ref{Hclass}), is also an eigenstate of angular momentum,
\beq
L\ketr{m,n}=\hbar(n-m) \ketr{m,n}.
\eeq
We further introduce the left radius-squared operator,
\beq
R^2_L \equiv X^2_L + Y^2_L = \theta(2\bld \bl+1),
\eeq
and note that the states (\ref{mn}) are simultaneously eigenstates of this operator and of $L$ (clearly $R^2_L$ and $L$ commute),
\beq
R^2_L\ketr{m,n} = \theta (2n+1)\ketr{m,n} \;\forall \;n.
\eeq
In this light it is clear that the ``minimal radius-squared'' state of the form (\ref{mn}) is that which has a zero state label in the left
sector ($m=0$) and is consequently the eigenstate of $R^2_L$ with the smallest eigenvalue,
\beq
\ketr{0,n}\equiv\ket 0 \bra n,\quad R^2_L\ketr{0,n} = \theta \ketr{0,n} \;\forall \;n.
\label{0n}
\eeq
One should take care not to interpret this as a statement about the physical size / distribution of the state $\ketr{0,n}$. We associate the left sector with position in our description, and consequently one should read the above equation as stating that such a state is localised
on the circumference a disk of area $\pi\theta$ about the origin (since the radius of such a disk is $\sqrt \theta$). The fact that the minimal area is not zero is simply a manifestation of the quantisation of space induced by the non-commutativity of co-ordinates. The states (\ref{0n}) have the further property that their right sector, characterised by $n$, simply labels their angular momentum,
\beq
L \ketr{0,n} = \hbar n \ketr{0,n}.
\eeq
We now have a clear physical picture for these states. As was previously done for the states $\ket 0 \bra v$ in (\ref{zv2}), we now define translations of the states (\ref{0n}),
\beqa
\ketr{z,n}&\equiv&T(z)\ket{0}\bra{n} \nl
&=& \ket z \bra n e^{\zb b - z b^\dagger},
\label{zn}
\eeqa
so that
\beq
\bl \ketr{z,n} = z \ketr{z,n}.
\eeq
We may thus interpret $z$ as the positional state label, i.e., such a state is then a position state in the sense of Section \ref{zalphasection1}. Since the bosonic states allow a resolution of the identity on $\hc$,
\beq
\sum_{n=0}^\infty \ket n \bra n = \mathbf{1}_c,
\eeq
it is clear from (\ref{identityzalphaproof}) and (\ref{identityzalpha}) that we may also resolve the identity on the quantum Hilbert space in terms of the states (\ref{zn}),
\beq
\frac{1}{\pi}\int d^2z \sum_{n=0}^\infty \ketr{z,n} \brar{z,n} = \mathbf{1}_q,
\label{identityzn}
\eeq
i.e., we have found another decomposition of the star product in (\ref{identityz}),
\beq
\sum_{n=0}^\infty \ketr{z,n}\brar{z,n} = \ketr{z}\star\brar{z}.
\label{starsum}
\eeq
As regards the requirement of positivity, it is clear that
\beqa
\overlapr{\phi}{z,n}\overlapr{z,n}{\phi} &=& \tr{\phi^\ddagger \ket z \bra n e^{\zb b - z b^\dagger}} \tr{[\ket z \bra n e^{\zb b - z b^\dagger}]^\ddagger \phi} \nl
&=& \matrel{n}{e^{\zb b - z b^\dagger} \phi^\ddagger}{z} \matrel{z}{\phi e^{z b^\dagger -\zb b } }{n} \nl
&=& | \matrel{n}{e^{\zb b - z b^\dagger} \phi^\ddagger}{z} |^2 \geq 0 \;\forall\;\phi.
\eeqa
Consequently we may again introduce a POVM corresponding to the states (\ref{zn})
\beq
\pi_{z,n} \equiv \frac{1}{\pi}\ketr{z,n}\brar{z,n},\quad \int d^2z \sum_{n=0}^\infty\;\pi_{z,n} = \mathbf{1}_q,
\label{pizn}
\eeq
in terms of which we may define another local probability distribution. Assuming the system is in a pure state $\ketr{\psi}$, this distribution in $z$ and $n$ is simply
\beq
\label{Pzn}
P(z,n)=\brar{\psi}\pi_{z,n}\ketr{\psi}=\frac{1}{\pi}\overlapr{\psi}{z,n}\overlapr{z,n}{\psi} = \frac{1}{\pi}\left| \overlapr{z,n}{\psi} \right|^2.
\eeq

Returning to the physical interpretation of the new state label $n$, we note that the states (\ref{zn}) are eigenstates of the translated left
radius-squared operator,
\beq
T(z)\,R^2_L\,T^\ddagger(z)\ketr{z,n} = \theta \ketr{z,n}.
\eeq
This statement also applies to the translated angular momentum operator,
\beq
T(z)\,L\,T^\ddagger(z)\ketr{z,n} = \hbar n \ketr{z,n}.
\eeq
The reasoning is as previously: we have translated the eigenstates  $\ket 0 \bra n$ of $L$ and $R^2_L$ from the origin to the point $z$, and thus we must also translate the operators themselves to satisfy the eigenvalue equation. Consequently the $n$ in $\ketr{z,n}$ refers to the angular momentum about the point $z$. At this level already we note that the notion of some ``additional structure'' is clearly present: if an object is localised at the point $z$, and it has an angular momentum about this point, then by necessity the object must have some sort of non-trivial structure (i.e., it cannot be a point particle).

\section{Relating the states $\ketr{z,v}$ and $\ketr{z,n}$}

Previously we considered translations of the form
\beq
\ketr{z,v}=T(z)\ket{0}\bra{v}.
\eeq
Clearly this state is simply a coherent state in the angular momentum sector (labeled by $v$), that is
\beq
\ketr{z,v} = e^{-|v|^2/2} \sum_{n=0}^\infty \frac{\vb^n}{\sqrt{n!}}\ketr{z,n}.
\label{basistrans1}
\eeq
It is, of course, possible to perform the inverse basis transformation as well:
\beq
\ketr{z,n} =  \frac{1}{\pi} \int d^2v\,e^{-|v|^2/2}\,\frac{v^n}{\sqrt{n!}} \ketr{z,v}.
\label{basistrans2}
\eeq

At this point it should be noted that the probability distributions (\ref{Pzn}) and (\ref{Pzv}) for a particular system are related. To see
this we define a generating functional
\beq
G(\lambda) = \frac{1}{\pi} \overlapr{\psi}{z,v} e^{\lambda \stackrel{\leftarrow}{\partial_{\bar{v}}}\stackrel{\rightarrow}{\partial_v}}
\overlapr{z,v}{\psi}.
\eeq
Next we note that $P(z,v) =\frac{1}{\pi}|\overlapr{z,v}{\psi}|^2=\frac{1}{\pi} | \matrel{z}{\psi e^{z b^\dagger-\zb b}}{v} |^2$ and
$P(z,n) = \frac{1}{\pi} | \matrel{z}{\psi e^{z b^\dagger-\zb b}}{n} |^2 $, which implies that
\beq
P(z,v) = G(\lambda)|_{\lambda=0}\quad\textnormal{and}\quad P(z,n)= \left.\frac{\partial_\lambda^n}{{n!}} G(\lambda)\right|_{                                                                                                                          \lambda=0;\;v=\vb=0}.
\eeq
Stated differently, we may obtain the distribution (\ref{Pzn}) from the distribution (\ref{Pzv}) through
\beq
P(z,n)=\left.\frac{\partial_{v}^{n} \partial_{\vb}^{n}}{{n!}} P(z,v)\right|_{v=\vb=0}.
\eeq
Note that although we may transform from the state $\ketr{z,n}$ to the state $\ketr{z,v}$ (and thus also from the probability amplitude $\overlapr{z,n}{\psi}$ to $\overlapr{z,v}{\psi}$), it is not possible to transform from the distribution $P(z,n)$ to the distribution $P(z,v)$ in general.\footnote{ The reasoning here is that although $\overlapr\psi {z,v} = e^{-|v|^2/2}\sum_{n=0}^\infty \vb^n \overlapr \psi {z,n} $ through (\ref{basistrans1}), we have in general that
$\left|\overlapr \psi {z,v}\right|^2 \neq e^{-|v|^2}\sum_{n=0}^\infty \left|\vb^n \overlapr \psi {z,n}\right|^2$.} This is due to the fact that the right sector of basis elements $|z,v)$ is specified by a complex variable, and the coherent states $\ket v$ form an over-complete basis on $\hc$. The right sector of basis states $|z,n)$, in contrast, is characterised by bosonic states $\ket n$ which simply form a complete (orthonormal) basis for $\hc$. This also explains why constraints arise for the $\ketr{z,v}$ basis, but not in the $\ketr{z,n}$ basis.\\

To get a further idea of the link between the states $\ketr{z,v}$ (\ref{zv2}) and $\ketr{z,n}$ (\ref{zn}), let us consider again that the
state $\ketr{z,n}$ represents an object with angular momentum $\hbar n$ about the point $z$. As stated earlier and as is clear from the basis transformation (\ref{basistrans1}), the state $\ketr{z,v}$ is simply a weighted sum of such angular momentum states. Compare this to standard quantum mechanics: there an angular momentum state is of the form $e^{i m \phi}$. If we sum over all values of $m$ for such a state, we obtain a precise localisation in the angle, $\phi$ (in terms of a Dirac delta function). To see the analogy with the states discussed here, it is again useful to refer back to the picture of the two-charge composite, where $z$ refers to the co-ordinate of one charge and $v$ to the relative co-ordinate between the charges. In this context $v$ (which is the label of the coherent state in $n$, i.e., of a weighted summation of angular momentum states -- see (\ref{basistrans1})) determines the orientation, i.e., also an angular localisation of the composite.

Let us further consider the expectation value of angular momentum in the basis (\ref{zv2}),
\beq
\matrelr{z,v}{L}{z,v}= \hbar \left(|z+v|^2-|z|^2\right).
\eeq
It is clear that if $v=0$, the average angular momentum of such a state is zero, which implies that for this choice the ``orbital'' and ``intrinsic'' angular momenta cancel each other out on average. It is also clear that for the untranslated state $\ketr{0,v} = \ket 0 \bra v$, we have
\beq
\langle L \rangle_{\ketr{0,v}}\equiv\matrelr{0,v}{L}{0,v}= \hbar |v|^2 \quad\textnormal{with}\quad \Delta L = \sqrt{\langle L^2 \rangle_{\ketr{0,v}}-\langle L \rangle_{\ketr{0,v}}^2}=\hbar|v|.
\eeq
Again considering the dual picture of two harmonically interacting charges in a magnetic field set out in Section \ref{class}, we see that
the extent of the composite is directly proportional to the average angular momentum about its point of localisation --- this is, of course, logical, since higher angular
momentum about one of the charges (i.e., about the co-ordinate $z=0$) would cause a stretching of the spring between them (i.e., an increase in the length of $v$). Note that $v$ is dimensionless here, thus the relative fluctuations $\Delta L / \langle L \rangle$ decrease for higher average angular momenta.

\section{Average energy}
\label{averageE}
Consider now that we may write the states (\ref{zn}) as\footnote{ The second line of this equation allows for a very simple proof of (\ref{identityz}) given (\ref{identityzn}).}
\beqa
\ketr{z,n} &=& \frac 1 {\sqrt{n!}} \ket z \bra z (b-z)^n \nl
&=& \frac 1 {\sqrt{n!}} \left(\diffpa{\zb}\right)^n \ketr{z} \nl
&=& \frac 1 {\sqrt{n!}} \left(\frac 1 {i\hbar} \sqrt{\frac \theta 2} P\right)^n \ketr z,
\label{ztozntrans}
\eeqa
where we made use of definition (\ref{braz}) of the states $\ketr z$. This allows for an easy calculation of the expectation value of the free particle energy in the $\ketr{z,n}$-basis, namely
\beqa
\langle H_{free}\rangle_{\ketr{z,n}}&=&\matrelr{z,n}{\frac{P^\ddagger P}{2m}}{z,n} \nl
&=& \frac{1}{2m} \left[-\left(\frac 1 {i\hbar}\right)^2\frac \theta 2 \right]^{n-(n+1)} \frac{(n+1)!}{n!} \overlapr{z,n+1}{z,n+1}\nl
&=& \frac{\hbar^2}{m\theta}(n+1).
\label{freeexpzn}
\eeqa
Further it is straight forward to verify that
\beqa
\langle(H_{free})^m\rangle_{\ketr{z,n}}&=& \left(\frac{\hbar^2}{m\theta}\right)^m \frac{(n+m)!}{n!}.
\eeqa
Consequently the fluctuations pertaining to (\ref{freeexpzn}) are simply
\beqa
\Delta H_{free}&=& \sqrt{\langle (H_{free})^2\rangle - \langle H_{free}\rangle^2}\nl
&=& \frac{\hbar^2}{m\theta}\sqrt{n+1},
\eeqa
which in turn implies that the relative fluctuations $\Delta H_{free} / \langle H_{free}\rangle \sim 1/\sqrt{n}$ for large $n$, and thus decrease for large average energies. As expected, (\ref{freeexpzn}) is independent of $z$ since the free particle Hamiltonian is translationally invariant. In this sense there is a degeneracy in that the expected energy would be the same for all states $\ketr{z,n}$ for a particular $n$, independent of the position $z$. One may view this in analogy to the quantum Hall system, where it costs no energy to translate the particle in the plane. Furthermore, in the quantum Hall system, the particle may be better localised by exciting higher Landau levels (see discussion in \cite{jain}). Better localisation, in turn, requires high momenta. Comparing this to the expression above, we note that $n$ may be viewed as being analogous to a label of Landau levels. To make this explicit, let us relate the spectrum of the Landau problem, $E_n = \hbar \omega_c (n+1/2)$ with $\omega_c = \frac{|e|B}{m}$, to (\ref{freeexpzn}). Equating the coefficients of $n$ yields
\beq
\theta = \frac \hbar {|e| B},
\eeq
which again demonstrates that the non-commutative parameter scales as the square of the magnetic length ${\ell} \equiv \sqrt{\hbar/{|e| B}}$. If we were to add a potential to the Hamiltonian, i.e.,
\beq
H = H_{free} + V,
\eeq
we note that the expectation value of the energy would become
\beq
\langle H \rangle_{\ketr{z,n}} = \frac{\hbar^2}{m\theta}(n+1) + \matrel z V z,
\eeq
since the potential is a function of the co-ordinates and is thus insensitive to the right sector. This implies that the potential would lift the aforementioned degeneracy brought about by translational invariance. Assuming that $\theta$ is small, the energy scales involved in the external potential would be much smaller than those set by $\hbar^2/m\theta$.

We conclude that, on average, it takes large amounts of energy (due to the energy scale set by $\hbar/m\theta$) to excite higher values of $n$ for a free particle. This behaviour is analogous to that of a composite with great rotational inertia.

\section{Some probability distributions}

In this section we shall discuss differences between position measurements which probe the right sector (in terms of a local position description) and those that are insensitive to right sector degrees of freedom (in terms of the non-local position description).

\subsection{Pure position measurements in terms of the non-local POVM (\ref{povm})}
Suppose we have a system prepared in a minimal uncertainty position state (in the sense of \cite{jopa}) of the form $\ketr \psi = \ketr{w} = \ket w \bra w$ (i.e., a state which is a translation of the state $\ket 0 \bra 0$ where we do not specify the right sector). The probability distribution of position in terms of the non-local formalism set out in Chapter \ref{ncformalism} is simply
\beqa
P(z) &=& \matrelr w {\pi_z} w \nl
&=& \frac 1 \pi \overlapr{w}{z} \star \overlapr z w \nl
&=& \frac 1 \pi e^{-|z-w|^2}.
\label{Pzch61}
\eeqa
As expected (due to the over-completeness of the states (\ref{braz})) we obtain a Gaussian decay (with the length scale set by $\theta$) for the probability of finding the system in another minimal uncertainty state $\ketr z$.\footnote{ Upon restoring the correct dimensionality in the exponent in (\ref{Pzch61}) we see that this distribution becomes a Dirac delta function in the commutative limit, as expected.}

It would be natural to ask what would happen if we do in fact specify the right sector of the state of the system, but perform the same measurement. To this end, suppose now that we prepare the system in the state $\ketr \psi = \left|w,n\right)=T(w)\left|0\right\rangle \left\langle n\right|$. Again the probability of measuring position $z$ is given in terms of the POVM (\ref{povm}),
\beqa
P(z)=\left(w,n\right|\pi_{z}\left|w,n\right)=\frac{1}{\pi}\overlapr{w,n}{z}\star\overlapr{z}{w,n}.
\eeqa
Now
\beqa
\overlapr{z}{w,n} & = &  \textnormal{tr}_{c}\big[\left\{ T(z)\left|0\right\rangle \left\langle 0\right|\right\} ^{\dagger}\left\{T(w)\left|0\right\rangle \left\langle n\right|\right\}\big]\nl
 & = & \textnormal{tr}_{c}\big[\left|0\right\rangle \left\langle 0\right|\left\{ T(w-z)\left|0\right\rangle \left\langle n\right|\right\} \big]\nl
 & = & \left\langle 0\right|\big\{ T(w-z)\left|0\right\rangle \left\langle n\right|\big\} \left|0\right\rangle \nl
 &=& \overlap{0}{w-z} \overlap{n}{z-w}\nl
 &=& e^{-|z-w|^{2}}\frac{(z-w)^{n}}{\sqrt{n!}},
\eeqa
where we made use of the properties (\ref{Tzphi}) and (\ref{Tzproperties}) of the translation operator. Consequently
\beqa
\overlapr{w,n}{z} & = & e^{-|z-w|^{2}}\frac{(\bar{z}-\bar{w})^{n}}{\sqrt{n!}},
\label{overlapwnz}
\eeqa
which implies that
\beqa
P(z) & = & \frac{1}{\pi}\left\{ e^{-|z-w|^{2}}\frac{(\bar{z}-\bar{w})^{n}}{\sqrt{n!}}\right\} e^{\stackrel{\leftarrow}{\partial_{\bar{z}}}\stackrel{\rightarrow}{\partial_z}}\left\{ e^{-|z-w|^{2}}\frac{(z-w)^{n}}{\sqrt{n!}}\right\} \nl
 & = & \frac{1}{\pi}\left.\frac{1}{n!}\partial_\alpha^{n}\partial_\beta^{n}\sum_{m=0}^{\infty}\frac{1}{m!}\left\{ e^{-(z-w-\alpha)(\bar{z}-\bar{w})}(z-w-\alpha)^{m}\right\} \left\{ e^{-(z-w)(\bar{z}-\bar{w}-\beta)}(\bar{z}-\bar{w}-\beta)^{m}\right\} \right|_{\alpha=\beta=0}\nl
 & = & \frac{1}{\pi}\left.\frac{1}{n!}\partial_\alpha^{n}\partial_\beta^{n}e^{-|z-w|^{2}+\alpha\beta}\right|_{\alpha=\beta=0}\nl
 & = & \frac{1}{\pi}e^{-|z-w|^{2}}.
\label{Pzch62}
\eeqa
We note that the probabilities (\ref{Pzch61}) and (\ref{Pzch62}) are identical. The reason for this is that the non-local POVM $\pi_z$ asks questions about position (i.e., the left sector) only, and does not probe the additional structure of the right sector. Indeed, this POVM cannot extract any information about the right sector exactly due to the fact that it provides a description where this information has been averaged out. In that regard the Gaussian decay above also makes sense, since the state $\ketr{w,n}$ is minimally localised at $w$ and the probability of finding it elsewhere again decays on a length scale of $\theta$. Since this type of measurement effectively sums over the contributions from the right sector (see (\ref{starsum})), it is to be expected that, even though we specified the right sector in the state $\ketr \psi = \ketr{w,n}$, the resulting distribution (\ref{Pzch62}) is independent of $n$. This demonstrates explicitly that a measurement of position alone (in the sense of \cite{jopa}) cannot yield full information about the state of the system.

\subsection{Position measurements in terms of the local POVM (\ref{pizn}) -- probing the right sector}

We again consider a system prepared in a state $\ketr \psi = \ketr{w} = \ket w \bra w$, where we have not specified details of the right sector before applying a translation. According to (\ref{Pzn}) the probability distribution in $z$ and $n$ is simply
\beqa
P(z,n)&=&\brar{w}\pi_{z,n}\ketr{w}\nl
&=&\frac{1}{\pi}\left| \overlapr{z,n}{w} \right|^2\nl
&=&\frac{1}{\pi} e^{-2|z-w|^2}\frac{|z-w|^{2n}}{n!}.
\eeqa
In the position sector there is again a Gaussian decay associated with minimal uncertainty position states in the probability distribution. This behaviour is of course generic to all position states, as is expected from the discussion in Chapter \ref{rightsectorchapter}. In addition to this, however, we note a dependence on $n$. In fact, the distribution above is essentially a Gaussian located on a ring, whose radius is determined by $n$. Since the POVM $\pi_{z,n}$ probes the probability of finding a state with angular momentum $\hbar n$ about the point $z$, this result is not surprising. One may even link this distribution to the standard position representation of the Landau problem in commutative quantum mechanics, where $n$ is the label of the Landau level. This again relates to the analogies drawn in Section \ref{averageE}.

Again it would be sensible to ask what happens if we carry out the same measurement on a system where we specify the right sector upon preparation, i.e., $\ketr \psi = \ketr{w,m} = T(w) \ket 0 \bra m$. We first note that
\beq
\overlapr{z,n}{w,m}=\matrel{m}{e^{(z-w)b^\dagger -(\bar z - \bar w)b}}{n} e^{-z\bar{z}/2-w\bar{w}/2+w\bar{z}},
\eeq
which implies that if $z=w$, these states are in fact orthogonal. The probability distribution for $z$ and $n$ is now simply
\beqa
P(z,n) = \frac 1 \pi \overlapr{w,m}{z,n}\overlapr{z,n}{w,m} = \frac 1 \pi e^{-|z-w|^2}\left|\matrel{m}{e^{(z-w)b^\dagger -(\bar z - \bar w)b}}{n}\right|^2.
\eeqa
Clearly this distribution goes like $\delta_{n,m}$ if $z=w$. This indicates again that translations mix up the local degrees of freedom specified by the right sector, as is seen in the term $\left|\matrel{m}{e^{(z-w)b^\dagger -(\bar z - \bar w)b}}{n}\right|^2$ above. Consequently it would only be possible to specify the right sector of basis states globally (i.e., independently of the positional state label) if, in addition to translations, we would apply a further unitary transformation to the right sector, whose function would be to cancel the right action of the translation operator. This will be discussed in more detail in the context of gauge theories in Chapter \ref{gaugechapter}.

\subsection{Some transition probabilities between states}

Let us find the transition amplitude between two states of the form (\ref{zn}),
\beq
\ketr{z_i,n_i}\equiv T(z_i) \ket 0 \bra{n_i},\quad i=1,2.
\eeq
Under the assumption of free propagation the time evolution operator is simply
\beq
{U}  =  e^{-\frac{it}{2m\hbar}P P^\ddagger}.
\eeq
We are now interested in a matrix element of the form $\matrelr{z_2,n_2}{U}{z_1,n_1}$. To compute this, we insert a complete\footnote{ The proof of completeness for these states is set out in Appendix \ref{momentumappendix}.} set of momentum states (\ref{psik}),
\beqa
\matrelr{z_2,n_2}{U}{z_1,n_1}&=& \int d^2k \,e^{-\frac{it}{2m\hbar}|k|^2}\overlapr{z_1,n_1}{\psi_k}\overlapr{\psi_k}{z_2,n_2}\nl
&=&  \sqrt{\frac{\theta}{2m\hbar^2}}\frac{1}{\sqrt{n_1 !}}\frac{1}{\sqrt{n_2 !}}\int d^2k \, \big\{e^{-[\frac{it\hbar+m\theta}{2m\hbar^2}]k\bar k} e^{\frac{i}{\hbar}\sqrt{\frac{\theta}{2}}[z_2-z_1]\bar k} e^{\frac{i}{\hbar}\sqrt{\frac{\theta}{2}}[\bar z _2-\bar z _1] k}\nl
& &\qquad\qquad\qquad\qquad\qquad\qquad\qquad[(-{i}/{\hbar})(\sqrt{{\theta}/{2}})\,k]^{n_1}
[({i}/{\hbar})(\sqrt{{\theta}/{2}})\,\bar k]^{n_2}\big\}, \nl
\eeqa
where we made use of the fact that $P\ketr{\psi_k}=k\ketr{\psi_k}$ (and its Hermitian conjugate) in the first line, and evaluated the overlaps explicitly in the second line. Upon identifying
\beq
\Delta \equiv z_2-z_1\quad\textnormal{and}\quad \bar \Delta \equiv \bar z _2 - \bar z _1
\eeq
and performing the Gaussian integral explicitly, the matrix element may be cast into a more user-friendly form
\beqa
\matrelr{z_2,n_2}{U}{z_1,n_1}&=& \frac{m\theta}{it\hbar+m\theta}\frac{1}{\sqrt{n_1 !}}\frac{1}{\sqrt{n_2 !}}\left(\frac{\partial}{\partial \Delta}\right)^{n_2} \left(-\frac{\partial}{\partial \bar \Delta}\right)^{n_1}e^{-\frac{m\theta}{it\hbar+m\theta}|\Delta|^2}.
\label{znpropagator}
\eeqa
(Note that the commutative limit is indeed well-defined here: the $\theta$ in the denominator of the prefactor $\frac{m\theta}{it\hbar+m\theta}$ is canceled by the $\frac 1 {2 \theta}$ which appears when we transform between dimensionless and dimensionful co-ordinates). Further defining
\beq
\Gamma \equiv \frac{m\theta}{it\hbar+m\theta},
\eeq
we note that (\ref{znpropagator}) may be rewritten in terms of generalised Laguerre polynomials,
\beq
\textnormal{L}^{a}_{n}\left(x\right)\equiv \frac{x^{-a}e^x}{n!} (\partial_x)^n\left(e^{-x}x^{n+a}\right),
\label{generalisedLaguerre}
\eeq
as
\beq
\matrelr{z_2,n_2}{U}{z_1,n_1}=\sqrt{\frac{n_1!}{n_2!}} \Gamma^{n_2+1}\Delta^{n_2-n_1}e^{-\Gamma|\Delta|^2}\textnormal{L}^{n_2-n_1}_{n_1}\left(\Gamma|\Delta|^2\right).
\label{znpropagator2}
\eeq
This expression is symmetric under the exchange $n_1\leftrightarrow n_2$, as is readily verified using properties of (\ref{generalisedLaguerre}). As expected, this simplifies precisely to the free path integral propagator found in \cite{sunandan} for the case where $n_1=n_2=0$, i.e., when $\ketr{z_i,0}=\ket{z_i}\bra{z_i}\equiv\ketr{z_i}$ --- see (\ref{braz}):
\beq
\matrelr{z_2}{U}{z_1}=\frac{m\theta}{it\hbar+m\theta}e^{-\frac{m\theta}{it\hbar+m\theta}|z_1-z_2|^2}.
\label{zzpropagator}
\eeq
Comparing this to the commutative free particle propagator, where the exponential is a pure phase, we note that non-commutative transition probability (i.e., the modulus squared of (\ref{zzpropagator})) has an actual exponential decay. This is again due to the over-completeness of basis elements $\ketr z \equiv \ket z \bra z$; indeed, in the commutative case we are considering the transition probability between two exactly localised, orthogonal states (i.e., Dirac delta functions), whereas the initial and final states in the non-commutative scenario are non-orthogonal coherent states. Clearly, even for $t=0$, there is a non-zero overlap between the two states, which is not the case in the commutative system. Consider also that two time-scales feature in the transition probability between $z_1$ and $z_2$ (i.e., the modulus squared of the overlap (\ref{zzpropagator})). For $t \gg m\theta/\hbar$ the Gaussian decay disappears, and we again approach the standard propagator of the commutative system. For very small time-scales, however, i.e., for $t\ll m\theta/\hbar$ the Gaussian decay (resulting from the over-completeness of the basis states) is dominant. We further note that, as discussed in \cite{cspathint,smailagic}, the non-commutative parameter induces an ultraviolet cutoff for the free particle. Again this may be understood as suppression of high momenta due to exclusion of small positional length-scales through non-commutativity.

Let us return to the general form (\ref{znpropagator2}) of the propagator. Suppose we consider an initial state with zero angular momentum about the point $z_1$, i.e., $\ketr{z_1,0}$. In this case, the Laguerre polynomial in (\ref{znpropagator2}) is simply $1$. If we now ask about the transition amplitude to the same point (i.e., $z_2 = z_1$), we note that transitions to higher angular momenta about this point, i.e., to $n_2 \neq 0$, are suppressed through the factorial prefactor. This is exactly what we expect, since it would cost energy to excite higher angular momenta.

\chapter{THE RIGHT SECTOR VIEWED AS GAUGE DEGREES OF FREEDOM}
\label{gaugechapter}

From the preceding discussions it is clear that it is possible to formulate non-commutative quantum mechanical position measurements in a local manner. As stated, the advantage of such a local formulation is that it makes the additional structure (degrees of freedom) involved in such measurements explicit. Although we discussed two specific bases for the right sector, there is no \emph{a priori} reason for a particular choice of basis. It would thus be natural to ask whether there exists a version of the theory that is insensitive to these specific choices of local (right sector) basis, thereby allowing for a local description that incorporates this additional structure in a more generic manner. In this section we shall formulate a gauge-invariant non-commutative Hamiltonian theory, and proceed to demonstrate that local transformations of the right sector (i.e., position-dependent transformations that probe the left sector but only transform the right sector degrees of freedom) may be absorbed as local gauge transformations in this context.

\section{A gauge-invariant formulation}

Consider a Hamiltonian in minimal coupling form,
\beq
H = \frac 1 {2m} (\vec P + e \vec A)^2,
\label{Hmincoupling}
\eeq
where $\vec A = (A_x, A_y)$ is the vector potential (gauge field). Note that $\vec A$ is a quantum operator here, and acts on the full quantum Hilbert space. Upon introducing the operators
\beq
D \equiv \frac 1 {\sqrt{2}} (P+ e \mathcal{A}) \quad \textnormal{and} \quad D^\ddagger \equiv \frac 1 {\sqrt{2}} (P^\ddagger+ e \mathcal{A}^\ddagger),
\label{Dgauge}
\eeq
where we define
\beq
\mathcal{A} \equiv A_x + i A_y \quad \textnormal{and} \quad \mathcal{A}^\ddagger \equiv A_x - i A_y
\label{Agauge}
\eeq
in analogy to the complex momenta $P$ and $P^\ddagger$ from (\ref{PPD}), we see that (\ref{Hmincoupling}) may be written as
\beq
H = \frac 1 {2m} (D^\ddagger D + D D^\ddagger).
\label{Hgauged}
\eeq
We proceed as usual with deriving the transformation rules for the gauge field. The time-independent Schr\"odinger equation reads
\beq
H \psi = E \psi.
\eeq
Suppose we now transform the wave-function by a unitary transformation, $U$ with $U^\ddagger = U^{-1}$, according to
\beq
\psi \rightarrow \psi ' \equiv U \psi,
\label{psiprime}
\eeq
and simultaneously transform the Hamiltonian (\ref{Hgauged}) according to
\beq
H \rightarrow H' \equiv U H U^\ddagger.
\label{Hprime}
\eeq
Clearly this yields a new eigenvalue equation,
\beq
H' \psi ' = E \psi ',
\eeq
i.e., $H$ and $H'$ are isospectral. It is a simple matter to verify that
\beq
H' = \frac 1 {2m}(D' D'^\ddagger + D'^\ddagger D')
\eeq
is gauge-invariant under the transformation rules
\beqa
D' = \frac 1 {\sqrt 2} (P + e\mathcal A ') \;\; &\textnormal {with}& \;\;\mathcal{A} '= U^\ddagger\mathcal A U - \frac 1 e [P,U^\ddagger]U, \;\;\textnormal{and} \nl
D'^\ddagger = \frac 1 {\sqrt 2} (P^\ddagger + e\mathcal A '^\ddagger) \;\; &\textnormal {with}& \;\; \mathcal{A} '^\ddagger = U^\ddagger \mathcal A ^\ddagger U - \frac 1 e [P^\ddagger,U^\ddagger]U,
\eeqa
i.e., that
\beq
U H[\mathcal{A} ' , \mathcal A '^\ddagger] U^\ddagger = H[\mathcal {A} , \mathcal {A} ^\ddagger].
\eeq
Note that the operators $D$ and $D^\ddagger$ are analogues of the covariant derivative from commutative gauge theories, and that they obey the transformation rules
\beqa
D[\mathcal{A}'] &=& U D[\mathcal{A}]U^\dagger \;\; \textnormal{and} \nl
D[\mathcal{A}'^\dagger] &=& U D[\mathcal{A^\dagger}]U^\dagger
\label{Dtransformations}
\eeqa
by construction.

Hereby we have derived the transformation rules for the gauge field. These rules are essentially identical to those that would be found in a commutative theory, except that the momenta act adjointly (i.e., as algebraic derivatives) in the non-commutative theory.\footnote{ In standard quantum mechanics, a pure gauge transformation of a Hamiltonian $H = \frac 1 {2m}(\vec P + \vec A)^2\;\rightarrow H'=UHU^\dagger$ and the corresponding wave function $\psi \rightarrow \psi' = U\psi$, is one which induces additional terms the gauge field (vector potential) $\vec A$ which are of the form $A_i = U(\partial_i U^\dagger)$. This simply implies that the curl of these additional terms is zero, i.e., no new magnetic field is introduced by the transformation. Consequently the two Hamiltonians represent the same physical situation, i.e., a pure gauge transformation introduces no new physics.} Under these transformation rules, the Hamiltonian (\ref{Hgauged}) thus displays a gauge symmetry under the gauge transformation set out in (\ref{psiprime}) and (\ref{Hprime}).

\section{A local transformation of the right sector seen as a gauge transformation}

Let us consider again a generic basis element of the form (\ref{Tzalpha}),
\beq
\ketr{z,\alpha}\equiv T(z) \left(\ket{0}\bra{\alpha}\right)=\ket{z}\bra{\alpha}e^{\zb b-z b^\dagger},
\eeq
where $\alpha$ may be a discrete (i.e., Fock basis) or continuous (i.e., coherent state basis) label for the right sector. Suppose we now introduce a new basis that is obtained by a local (in the sense that it probes information regarding position, i.e., about the left sector) similarity transformation which transforms the right sector and whose inverse is well-defined. Such a transformation would necessarily have left- and right acting parts, and would be a function of  $\bl$, $\br$ and $\brd$,
\beq
\ketr{z,\beta} = S(\bl, \br,\brd) \ketr{z,\alpha} = S(z,\br,\brd) \ketr{z,\alpha} \;\;\Leftrightarrow\;\;\ketr{z,\alpha} = S^{-1}\ketr{z,\beta}.
\label{basistransgauge}
\eeq
One may, of course,  also view this basis transformation in terms of expansion co-efficients,
\beqa
\ketr{z,\beta} &=& \int\, d^2 z' \sum_\eta \ketr{z',\eta} \matrelr{z',\eta} {S(\bl, \br,\brd)} {z,\alpha} \nl
&=& \int\, d^2 z' \sum_\eta C({z, z',\eta,\alpha}) \ketr{z',\eta} \quad\textnormal{with}\quad C({z, z',\eta,\alpha}) \in \mathbb{C}.
\eeqa
We have thus established a $1-1$ invertible mapping between two sets of ``local'' basis states,
\beq
\zeta_\alpha (z) \equiv \textnormal{span}_\alpha\left\{\ketr{z,\alpha}\right\} \quad \textnormal{and}\quad
\zeta_\beta (z) \equiv \textnormal{span}_\beta\left\{\ketr{z,\beta}\right\}
\eeq
so that
\beqa
&\forall \;\ketr{z,\beta} \;\in \;\zeta_\beta (z) \; \exists \;\ketr{z,\alpha}\;\in \;\zeta_\alpha (z):& \ketr{z,\beta}= S\ketr{z,\alpha} \nl
&\textnormal{and}& \nl
&\forall \;\ketr{z,\alpha} \;\in \;\zeta_\beta (z) \;\exists \;\ketr{z,\beta}\;\in \;\zeta_\beta (z):& \ketr{z,\alpha}=S^{-1}\ketr{z,\beta}.
\eeqa
Consequently, if the label $\alpha$ is discrete (continuous) then $\beta$ must also be discrete (continuous) and vice-versa, since there is not a $1-1$ correspondence of elements between a complete and an over-complete basis.

Let us return to the Hamiltonian (\ref{Hgauged}), the action of which may now be written in this new basis as
\beqa
\matrelr{z,\beta}{H}{\psi} &=& \matrelr{z,\alpha}{S^{\ddagger}H}{\psi}\nl
&=& \matrelr{z,\alpha}{S^\ddagger H(S^\ddagger)^{-1}S^\ddagger}{\psi}\nl
&=& \matrelr{z,\alpha}{ H'}{\psi'},
\eeqa
with $H'\equiv S^{\ddagger}H(S^{\ddagger})^{-1}$ and $\ketr{\psi'}\equiv S^{\ddagger}\ketr{\psi}$. Suppose that the transformation $S$ (\ref{basistransgauge}) (and thus its Hermitian conjugate on $\hq$, $S^\ddagger$) is indeed unitary. In this case the connection to the discussion from the previous section is clear: if the action of the Hamiltonian (\ref{Hgauged}) is expressed in a particular local basis, a unitary transformation of the right sector of this basis takes precisely the form of a pure gauge transformation performed on $H$ and $\psi$. For the more general case where $S$ is a pure similarity transformation, i.e., it is invertible but not unitary, the new Hamiltonian $H'$ may be non-Hermitian; however, the arguments about spectra still hold.\footnote{ Note that the discussion presented here applies to \emph{any} similarity transformation $S$ in (\ref{basistransgauge}), not only those that transform the right sector only. However, if we wish the new basis to be a ``position basis'' in the sense of Chapter \ref{rightsectorchapter}, then this transformation must necessarily be one that only transforms the right sector of the basis elements. We shall restrict our discussions to such transformations.} Note that the local transformation of the right sector acts on an infinite dimensional space (since there are infinitely many state labels for the right sector).

We thus have a formulation of the non-commutative theory which allows for total arbitrariness of local (right sector) choices of basis in the context of gauge invariance.

\section{Adding dynamics for the gauge field -- some cautious speculations}

Thus far we have only considered the minimal coupling form of the Hamiltonian (\ref{Hmincoupling}). In this form, the gauge field is handled as a background field in that there are no dynamics ascribed to it. If we wished to make the gauge field dynamical, this would involve adding a term which is a function of the field to the Hamiltonian. A further analysis should probably best be continued on the level of the action, as is usually done in the context of standard gauge theories. In this setting it should be possible to construct a term analogous to the usual $\textnormal{tr} [F_{\mu\nu}F^{\mu\nu}]$ term, which possesses the desired invariance properties. Suppose we introduce an object\footnote{ This object is exactly analogous to that from standard quantum mechanical gauge theories, namely \\$F_{\mu,\nu} = \partial_\mu A_\nu - \partial_\nu A_\mu + e [A_\mu,A_\nu]$. In standard electromagnetism, which is an Abelian gauge theory, this is simply the anti-symmetric electromagnetic field tensor, $F_{\mu,\nu} = \partial_\mu A_\nu - \partial_\nu A_\mu$.}
\beqa
F &\equiv& \frac 1 e[ D, D^\ddagger ] \nl
&=& [ P,\mathcal A ^\ddagger ] - [ P^\ddagger, \mathcal A ] + e [ \mathcal A , \mathcal A ^\ddagger ].
\label{Fgauge}
\eeqa
From (\ref{Dtransformations}) it is clear that
\beq
F' \equiv F[A'] = U F[A] U^\ddagger,
\eeq
as required. Although we shall not investigate this idea in great depth here, we shall mention a few ideas for possible future work. In principle (\ref{Fgauge}) could be used to build the gauge-invariant objects mentioned above. One may speculate that since such terms govern the spectrum of the right sector degrees of freedom, it may be possible to restrict which of these degrees of freedom are accessible to the system at a certain energy / temperature. At low energies, this may imply that the system can only access a finite number of these degrees of freedom (of which there are infinitely many, since the right sector is infinite-dimensional). It is not clear which gauge symmetry would be associated with this restriction, but this may be one way to construct a low-energy effective theory for the right sector which essentially masks certain degrees of freedom and thus reproduces smaller gauge symmetries.

\chapter{DISCUSSION AND CONCLUSIONS}

We have argued that, in contrast to commutative quantum mechanics, the notion of position and its measurement in non-commutative space cannot yield complete information regarding the quantum state of a particle. Indeed, the introduction of additional structure is unavoidable in any position representation of non-commutative quantum mechanics. In the non-local formulation set out in \cite{jopa} this additional structure is encoded in higher order positional derivatives, whereas the local descriptions set out here require the introduction of additional degrees of freedom that label the right sector of basis states. We argued that these additional degrees of freedom appear naturally and unavoidably in any local position description of non-commutative quantum mechanics. This stands in contrast to state labels such as spin from standard quantum mechanics, which need to be added by hand. We demonstrated that it is entirely sensible to define such local position states, and that the right sector is in essence arbitrary and unnecessary for describing the position of a particle only. In order to gain insight about the physical meaning of the right sector, two specific choices of basis were investigated.

As set out in \cite{chris}, constants of motion found from the path integral representation in \cite{sunandan} show that already in a non-local, unconstrained description of non-commutative quantum mechanics there are hints at extended objects in the theory. Motivated by these findings, we demonstrated that for one particular choice of local basis with a continuous label for the right sector, one way to view aforementioned additional structure may be in terms of physical extent. It was demonstrated explicitly in the classical picture that the energy contains correction terms proportional to the non-commutative parameter. These corrections could also be cast in a local or non-local form. In the local formulation the Lagrangian of a free particle coincides precisely with that of two oppositely charged particles coupled by a harmonic potential and moving in a strong magnetic field. Using these results as a primer, we proceeded to show that an interpretation of non-commutative quantum mechanics in terms of extended objects with additional structure is indeed a natural one by considering representations of the angular momentum operator and various Hamiltonians as well as the corresponding eigenfunctions in this basis. Furthermore, constraints were shown to arise in this basis, and it was suggested that eigenfunctions of quantum operators may be obtained on a full function space, and that the physical eigenfunctions may then be selected as those that satisfy these constraints.

A further local choice of basis was suggested, namely one where the additional degrees of freedom (in this case discrete state labels) describe angular momentum about the positional state label of the basis elements. Analogies to the quantum Hall system were pointed out, and it was argued that the non-local POVM from \cite{jopa} cannot resolve information about the right sector of states of a non-commutative quantum system. The latter point demonstrates clearly that a measurement of position alone cannot yield sufficient information to specify the state of a non-commutative quantum system completely. These basis states were also used to find a local form of the non-commutative free particle propagator which describes the transition amplitude between states localised at certain points and with particular angular momenta about those points. Furthermore relations between the two local bases were pointed out.

It was also observed that all positional descriptions of non-commutative quantum mechanics, local or non-local, are equivalent in that the notion of additional structure is simply encoded in different ways, but information is not lost when one description is contrasted with another. In this sense the non-local formulation could be viewed as an effective description in that it averages over the additional degrees of freedom that appear explicitly in the local formulations. The local description of non-commutative quantum mechanics in terms of a constrained system may also offer an interesting new perspective.

Lastly it was argued that, if we were to insist that the local choice of basis for the right sector should be physically irrelevant, it may be natural to think of the additional degrees of freedom as gauge degrees of freedom. After formulating a gauge-invariant version of the non-commutative theory on the level of a Hamiltonian in minimal coupling form, we demonstrated that local transformations of the right sector may be absorbed into gauge transformations of this Hamiltonian and its eigenfunctions. First steps toward ascribing dynamics to the gauge field were suggested, and it was speculated that energetic considerations could possibly be employed to impose a restriction on which gauge degrees of freedom are accessible to the system, thereby altering the symmetry in the theory. This particular point may merit further investigation in the context of a Lagrangian formulation in the future.


\appendix

\chapter{Inclusion of a third co-ordinate}
\label{thirdcoord}
Suppose we have a three dimensional non-commutative Heisenberg algebra
\beqa
\left[x_i,x_j\right] & = & i \theta_{i,j} \quad\textnormal{with}\quad i,j=1,2,3,\nl
\left[x_{i} , p_{j} \right] & = & i \hbar \delta_{i,j},\nl
\left[ p_i,p_j \right]&=&0.
\eeqa
In this case it is clear that $\theta_{i,j}$ must be a completely antisymmetric matrix with real entries (otherwise the position operators cannot be Hermitian). Antisymmetry of a matrix implies that its eigenvalues come in pairs, $\{ \lambda_i,-\lambda_i\}$, which in turn implies that if the dimension of the matrix is odd, one eigenvalue must be zero. It is thus possible to perform a series of transformations on the co-ordinates that will diagonalise $\theta_{i,j}$ so that we have two non-commutating co-ordinates, each of which commutes with the third one.

Naturally this procedure breaks rotational symmetry of this framework, since the choice of co-ordinate orientation is not arbitrary. The only way to remedy this is to insist that $\theta_{i,j}$ should transform as a tensor. This point of view is naturally incompatible with the assumption that the non-commutative parameter should be a constant which is equal in all reference frames. The tensorial transformation properties of non-commuting co-ordinates and the restoration of rotational symmetry in higher dimensions are discussed extensively in \cite{rotation1,rotation2} in the setting of twist deformations and Hopf algebraic techniques. Since this thesis deals with a two-dimensional framework where these complications do not arise, this matter is not explored further here.\footnote{ It should be noted, however, that there is a required modification to the angular momentum operator (generator of rotations), even in a two-dimensional setting. This was discussed extensively in \cite{jopa}, and is also used in Section \ref{ang}.}

\newpage
\chapter{Proof of equation (\ref{rootpiz})}
\label{POVMsquaredProof}

Recall that we defined the POVM (\ref{povm}) as

\beq
\pi_z=\frac{1}{\pi}|z) \star (z|=\frac{1}{\pi}\ketr z e^{\stackrel{\leftarrow}{\partial_{\bar{z}}}\stackrel{\rightarrow}{\partial_z}}\brar z.
\eeq

Let us now consider the following matrix element
\def\nt{\tilde n}
\def\mt{\tilde m}

\beqa
\matrelr{\nt,\mt}{\pi_z}{n,m}&=&\frac{1}{\pi}\overlapr{\nt,\mt}{z}\star\overlapr{z}{n,m}\nl
&=&\frac{1}{\pi}\left(\overlap{z}{\mt}\overlap{\nt}{z}\right)\star\left(\overlap{z}{n}\overlap{m}{z}\right)\nl
&=&\frac{1}{\pi}\left(e^{-z\zb}\frac{\zb^{\mt}}{\sqrt{\mt !}} \frac{z^{\nt}}{\sqrt{\nt !}}\right)\star\left(e^{-z\zb}\frac{\zb^{n}}{\sqrt{n !}} \frac{z^{m}}{\sqrt{m !}}\right)\nl
&=&\frac{1}{\pi}\frac{z^{\nt}}{\sqrt{\nt !}}\frac{\zb^{n}}{\sqrt{n !}}\underbrace{\left(e^{-z\zb}\frac{\zb^{\mt}}{\sqrt{\mt !}} \right)e^{\stackrel{\leftarrow}{\partial_{\bar{z}}}\stackrel{\rightarrow}{\partial_z}}\left(e^{-z\zb} \frac{z^{m}}{\sqrt{m !}}\right)},\\
&&\qquad\qquad\qquad\qquad\qquad\qquad\equiv f(z)\nonumber
\label{pizproof1}
\eeqa

where we made use of the fact that $\overlap{n}{z}=e^{-|z|^2/2}\frac{z^{n}}{\sqrt{n !}}$. Next we note that

\beqa
f(z) &=& \frac{1}{\sqrt{\mt!}} \frac{1}{\sqrt{m!}}\sum_{k=0}^\infty \frac{1}{k!}  \left(\partial_{\zb}^k [\zb^{\mt}e^{-z\zb}] \right) \left(\partial_{z}^k[z^{m}e^{-z\zb}]\right)\nl
&=&\frac{1}{\sqrt{\mt!}} \frac{1}{\sqrt{m!}}\partial_{\lambda_1}^{\mt} \partial_{\lambda_2}^{m}\left.\sum_{k=0}^\infty \frac{1}{k!} \left(\partial_{\zb}^k [e^{-z\zb+\lambda_1\zb}] \right) \left(\partial_{z}^k[e^{-z\zb+\lambda_2 z}]\right)\right|_{\lambda_1=\lambda_2=0}\nl
&=&\frac{1}{\sqrt{\mt!}} \frac{1}{\sqrt{m!}}\partial_{\lambda_1}^{\mt} \partial_{\lambda_2}^{m}\left. e^{(\lambda_1 -z)(\lambda_2 -\zb)-2z\zb+\lambda_1 \zb + \lambda_2 z} \right|_{\lambda_1=\lambda_2=0}\nl
&=&\frac{1}{\sqrt{\mt!}} \frac{1}{\sqrt{m!}}\partial_{\lambda_1}^{\mt} \partial_{\lambda_2}^{m}\left. e^{\lambda_1 \lambda_2 - z\zb} \right|_{\lambda_1=\lambda_2=0}\nl
&=&e^{-z\zb}\frac{m!}{(\sqrt{m!})^2}\delta_{m,\mt}.
\eeqa

In the last step we simply note that if we expand the exponential, only terms of equal order in $\lambda_1$ and $\lambda_2$ will survive when we impose $\lambda_1=\lambda_2=0$, which produces the Kronecker delta. Inserting this result into (\ref{pizproof1}) we obtain

\beqa
\matrelr{\nt,\mt}{\pi_z}{n,m}&=&\frac{1}{\pi}\frac{z^{\nt}}{\sqrt{\nt!}}\frac{\zb^{n}}{\sqrt{n!}}e^{-z\zb}\delta_{m,\mt}\nl
&=& \frac{1}{\pi} \overlap{\nt}{z} \overlap{z}{n}\delta_{m,\mt}.
\label{pizproof2}
\eeqa

Next we define the operator

\beq
P_z \equiv \sum_{k=0}^\infty \ketr{z,k} \brar{z,k} \quad \textnormal{with} \quad \ketr{z,k}\equiv\ket z \bra k,
\eeq

and consider the following matrix element,

\beqa
\matrelr{\nt,\mt}{P_z}{n,m}&=& \sum_{k=0}^\infty \tr{\ket \mt \overlap{\nt}{z}\bra k} \tr{\ket k \overlap{z}{n}\bra m} \nl
&=& \sum_{k=0}^\infty \overlap{\nt}{z}\overlap{k}{\mt}\overlap{z}{n}\overlap{m}{k}\nl
&=& \overlap{\nt}{z} \overlap{z}{n}\delta_{m,\mt}.
\label{pizproof3}
\eeqa

Comparing (\ref{pizproof2}) and (\ref{pizproof3}), we note that

\beq
\pi_z = \frac{1}{\pi} P_z.
\eeq

Further, it is trivial to check that $P_z^2=P_z$. This simply implies that $\pi_z^2=\frac{1}{\pi^2}P_z^2=\frac{1}{\pi}\pi_z$, and consequently

\beq
\pi_z^{1/2}=\sqrt{\pi} \pi_z,
\eeq

which proves the result. \begin{flushright}{$\Box$}\end{flushright}

\newpage
\chapter{The path integral action}
\label{PathIntAct}

We follow here the discussion of \cite{klauder} where the propagator (matrix element of the time evolution operator) is found as a coherent state path integral.

Suppose we consider such a matrix element in the basis (\ref{zv2}),

\beq
\matrelr{z'',v''}{e^{-\frac{i}{\hbar}T H}}{z',v'},
\label{propagator1}
\eeq
where $U\equiv e^{-\frac{i}{\hbar}T H}$ is the unitary time evolution operator, $T=t''-t'$ represents the time interval of propagation and $H$ is the Hamiltonian of the system. We proceed by ``slicing'' the time interval into $N+1$ subintervals (the reason for the $+1$ will become evident shortly) of width $\epsilon = T/(N+1)$ so that $(z_{N+1},v_{N+1}) \equiv (z'',v'')$, $(z_{0},v_{0}) \equiv (z',v')$, and the $N$ co-ordinates $\left\{(z_{k},v_{k}),\; k=1:N\right\}$ represent $z$ and $v$ at time-subinterval $k$ of the time-sliced path. Naturally we may resolve the identity (\ref{identityzv}) at each of these points,

\beq
\int \ketr{z_k,v_k} \brar{z_k,v_k} d\mu_k = \mathbf{1}_q,\quad\textnormal{with}\quad d\mu_k \equiv \frac{1}{\pi^2}\,d^2 z_k \, d^2 v_k.
\label{propagator2}
\eeq

In order to compute the path integral, we introduce the regularised Hamiltonian

\beq
H_\epsilon = \frac{H}{1+\epsilon H^2} \quad \textnormal{with}\;\epsilon > 0,
\eeq
which has the property that $\lim_{\epsilon\rightarrow0}H_\epsilon = H$ (where the limits $\lim_{\epsilon\rightarrow0}$ and $\lim_{N\rightarrow\infty}$ are of course interchangeable). Next we write the time evolution operator as the limit

\beq
e^{-\frac{i}{\hbar}T H} = \lim_{N\rightarrow\infty} \left[ 1-\frac{i}{\hbar}\frac{T}{N+1}H_{T/(N+1)} \right]^{N+1}.
\eeq
We substitute this into (\ref{propagator1}) and subsequently insert the identity (\ref{propagator2}) at each point labeled by $k=1:N$, which yields
\beqa
& & \matrelr{z'',v''}{e^{-\frac{i}{\hbar}T H}}{z',v'} \nl
& &=\quad \lim_{N\rightarrow\infty}\matrelr{z'',v''}{\left[ 1-\frac{i}{\hbar}\epsilon H_{\epsilon} \right]^{N+1}}{z',v'} \nl
& &=\quad \lim_{N\rightarrow\infty} \int ... \int \;\prod_{k=0}^N \;\matrelr{z_{k+1},v_{k+1}}{\left[ 1-\frac{i}{\hbar}\epsilon H_{\epsilon} \right]^{N+1}}{z_{k},v_{k}} \;\prod_{k=1}^N d\mu_k \nl
& &=\quad \lim_{N\rightarrow\infty} \int ... \int \;\prod_{k=0}^N \; \overlapr{z_{k+1},v_{k+1}}{z_{k},v_{k}}\left[ 1-\frac{i}{\hbar}\epsilon \frac{\matrelr{z_{k+1},v_{k+1}}{H_\epsilon}{z_{k},v_{k}}}{\overlapr{z_{k+1},v_{k+1}}{z_{k},v_{k}}} \right]\;\prod_{k=1}^N d\mu_k\nl
\label{propagator3}
\eeqa
Proceeding to define
\beq
H_\epsilon(z_{k'},v_{k'};z_{k},v_{k})\equiv\frac{\matrelr{z_{k'},v_{k'}}{H_\epsilon}{z_{k},v_{k}}}{\overlapr{z_{k'},v_{k'}}{z_{k},v_{k}}},
\eeq
and assuming that the integrals exist, we rewrite (\ref{propagator3}) as
\beq
\matrelr{z'',v''}{e^{-\frac{i}{\hbar}T H}}{z',v'} = \lim_{N\rightarrow\infty} \int ... \int \;\prod_{k=0}^N \;\overlapr{z_{k+1},v_{k+1}}{z_{k},v_{k}}e^{-\frac{i}{\hbar}H_\epsilon(z_{k+1},v_{k+1};z_{k},v_{k})}\;\prod_{k=1}^N d\mu_k.
\label{propagator4}
\eeq
As $\epsilon\rightarrow 0$ we may view the set of points $(z_k,v_k),\;k=0:N+1$ as defining the limit of the functions $(z[t],v[t]),\;t\in[t',t'']$. If we wish to interchange the operations of integration and $\lim_{N\rightarrow\infty}$ in (\ref{propagator4}), however, it is necessary to assume that the integrand takes the form where aforementioned functions are continuous and differentiable paths $(z[t],v[t])$ in $z-v$ parameter space. For notational simplicity we shall denote $(z_t,v_t)\equiv(z[t],v[t])$.

Next we note that in the small $\epsilon$ limit we have
\beqa
\overlapr{z_{k+1},v_{k+1}}{z_{k},v_{k}} &=& 1-\brar{z_{k+1},v_{k+1}}\left\{ \ketr{z_{k+1},v_{k+1}}-\ketr{z_{k},v_{k}} \right\}\nl
&\cong& e^{-\brar{z_{k+1},v_{k+1}}\left\{ \ketr{z_{k+1},v_{k+1}}-\ketr{z_{k},v_{k}} \right\}}.
\eeqa
Introducing the notation
\beq
H(z,v)\equiv H(z,v;z,v)=\matrelr{z,v}{H}{z,v}
\eeq
and the differential of the state $\ketr{z_t,v_t}$,
\beq
d\ketr{z_t,v_t}\equiv\ketr{z_t+dz_t,v_t+dv_t}-\ketr{z_t,v_t},
\eeq
we see that, under the above assumptions, the integrand in (\ref{propagator4}) takes the form
\beq
\textnormal{exp}\left[-\int_{z',v'}^{z'',v''}\brar{z_t,v_t}\left\{d\ketr{z_t,v_t}\right\}-\frac{i}{\hbar}\int_{t'}^{t''}H(z_t,v_t)dt \right].
\eeq

Lastly we introduce the time derivative of a state $\ketr{z_t,v_t}$ as
\beq
\dot{\overbrace{\ketr{z_t,v_t}}}\equiv \frac{d}{dt} \ketr{z_t,v_t},
\eeq
and the measure that represents integration over all paths in $z-v$ parameter space,
\beq
[\mathcal{D}\mu]\equiv\lim_{N\rightarrow\infty}\prod_{k=1}^N d\mu_k.
\eeq
This allows us to write
\beq
\int[\mathcal{D}\mu]\; \textnormal{exp}\left[ \frac{i}{\hbar} \int_{t'}^{t''}dt\, \left\{i\hbar (z_t,v_t \dot{\overbrace{\ketr{z_t,v_t}}} -H(z_t,v_t)\right\} \right],
\eeq
i.e., we may identify the path integral action as
\beqa
S &=& \int_{t'}^{t''}dt\, \left\{i\hbar (z_t,v_t \dot{\overbrace{\ketr{z_t,v_t}}} -H(z_t,v_t)\right\}\nl
&=& \int_{t'}^{t''}dt\, \matrelr{z,v}{i\hbar \frac{d}{dt} - H}{z,v}.
\eeqa

\chapter{Momentum eigenstates (\ref{psik}) as a complete basis for $\hq$}
\label{momentumappendix}

We recall the form of these states,
\beq
\ketr{\psi_{k}} =\aa e^{\ac(\bar{k}b+k\bd)}= \aa e^{-\ab |k|^{2}}e^{\ac k\bd} e^{\ac \bar{k}b}.
\eeq
First we show that these states are orthogonal:
\beqa
\overlapr{\psi_{k'}}{\psi_k} &=& \frac{\theta}{2\pi\hbar^2} \tr{[ e^{\ac(\bar{k}'b+k'\bd)}]^\ddagger[ e^{\ac(\bar{k}b+k\bd)}]}\nl
&=& \frac{\theta}{2\pi^2\hbar^2} \int d^2z\; \bra z  e^{-\ac(\bar{k}'b+k'\bd)} e^{\ac(\bar{k}b+k\bd)} \ket z \nl
&=& \frac{\theta}{2\pi^2\hbar^2} e^{-\frac{\theta}{4\hbar^2}(|k|^2+|k'|^2)-\frac{\theta}{2\hbar} \bar k ' k} \int d^2z\; e^{\ac \zb(k - k')} e^{\ac z(\bar k -\bar k ')}\nl
&=& \frac{1}{\hbar^2} e^{-\frac{\theta}{4\hbar^2}(|k|^2+|k'|^2)-\frac{\theta}{2\hbar} \bar k ' k}\,\delta (k - k').
\eeqa
To show that we may resolve the identity on $\hq$ in terms of these states, consider again the overlap (\ref{overlapz}) of two states of the form (\ref{braz}); for $\ketr{z_i}=\ket{z_i}\bra{z_i}$, $i=1,2$, we had
\beq
\overlapr{z_1}{z_2} = e^{-|z_1-z_2|^2}.
\eeq
Now, since
\beq
\overlapr{z_i}{\psi_k}=\aa e^{-\ab |k|^{2}}e^{\ac(\bar{k}z_i+k\zb_i)},
\eeq
we note that
\beqa
\int d^2k\,\overlapr{z_1}{\psi_k}\overlapr{\psi_k}{z_2} = e^{-|z_1-z_2|^2},
\eeqa
as is readily verified through explicit evaluation of the Gaussian integrals. We conclude that
\beq
\int d^2k\, \ketr{\psi_k}\brar{\psi_k}=\mathbf{1}_q
\eeq
is a resolution of the identity on $\hq$, i.e., the states (\ref{psik}) provide a complete basis for this space.

\specialhead{BIBLIOGRAPHY}


\markboth{}{}

\begin{thebibliography}{xx}
\bibitem{snyder} H. S. Snyder, Physical Review {\bf 71} (1947) 38.
\bibitem{dopplicher} S. Doplicher, K. Fredenhagen and J. E. Roberts, Commun. Math. Phys. {\bf 172} (1995) 187.
\bibitem{susskind} D. Bigatti and L. Susskind, Phys. Rev. D {\bf 62} (2000) 066004.
\bibitem{seiberg} N. Seiberg, ``Emergent space time'', arXiv:hep-th/0601234.
\bibitem{douglas} M. R. Douglas and N. A. Nekrasov, Rev. Mod. Phys. {\bf73} (2001) 97.
\bibitem{connes} A. Connes, M. R. Douglas, and A. Schwarz, J. High Energy Phys. {\bf02} 003-1–34; hep-th/9711162.
\bibitem{jain} J.K. Jain, {\it Composite Fermions} (Cambridge University Press, Cambridge, 2007).
\bibitem{thermodynamics} F.G. Scholtz and J. Govaerts, Jnl. Phys. A: Math. Theor. {\bf 41} (2008) 505003.
\bibitem{ncgeometry} R. Banerjee, B. Chakraborty, S. Ghosh, P. Mukherjee and S. Samanta, Found Phys {\bf 39} (2009) 1297–1345.
\bibitem{jopa} F.G. Scholtz, L. Gouba, A. Hafver and C.M. Rohwer, Jnl. Phys. A: Math. Theor. {\bf 42} (2009) 175303.
\bibitem{sakurai} J.J. Sakurai, {\it Modern Quantum Mechanics - Revised Edition} (Addison-Wesley Publishing Company, 1994).
\bibitem{bergou} J. A. Bergou, Jnl. Phys. Conf. Series {\bf 84} (2007) 012001.
\bibitem{neumann} J. von Neumann, {\it Mathematical Foundations of Quantum Mechanics} (Princeton: Princeton University Press, 1955).
\bibitem{stonevonneumann} G.W. Mackey, {\it The Theory of Unitary Group Representations} (The University of Chicago Press, 1976).
\bibitem{prosen} T. Prosen, New Journal of Physics {\bf 10} (2008) 043025.
\bibitem{preskill} J. Preskill, Lecture Note for Physics: Quantum Information and Computation, \emph{http://theory.caltech.edu/people/preskill}.
\bibitem{davies} E.B. Davies, \emph{Quantum Theory of Open Systems} (Academic Press, 1976).
\bibitem{torus} J. Govaerts and F.G. Scholtz, Jnl. Phys. A: Math. Theor. {\bf 40} (2007) 12415.
\bibitem{ncwell} F.G. Scholtz, B. Chakraborty, J. Govaerts and S. Vaidya, Jnl. Phys. A: Math. Theor. {\bf 40} (2007) 14581.
\bibitem{klauder} J. R. Klauder and B. Skagerstam, {\it Coherent States: Applications in Physics and Mathematical Physics} (World Scientific, Singapore, 1985).
\bibitem{holevo} A.S. Holevo, {\it Probabilistic and Statistical Aspects of Quantum Theory} (North-Holland Publishing Company, Amsterdam, 1982) p79.
\bibitem{sunandan} S. Gangopadhyay and F. G. Scholtz, Phys. Rev. Lett. {\bf 102} (2009) 241602.
\bibitem{chris} C.M. Rohwer, K.G. Zloshchastiev, L. Gouba and F.G. Scholtz, Jnl. Phys. A: Math. Theor. \textbf{43} (2010) 345302.
\bibitem{cspathint} S. Gangopadhyay and F.G. Scholtz, ``Free particle on noncommutative plane --– a coherent state path integral approach'', arXiv:0812.3474 (2008).
\bibitem{smailagic} A. Smailagic and E. Spallucci, Jnl. Phys. A: Math. Theor. {\bf 36} (2003) L467.
\bibitem{rotation1} P. G. Castro, B. Chakraborty, and F. Toppan, J. Math. Phys. {\bf49} (2008) 082106.
\bibitem{rotation2} B. Chakraborty, Z. Kuznetsova and F. Toppan, ``Twist Deformation of
Rotationally Invariant Quantum Mechanics'', arXiv:1002.1019v1 [hep-th] (2010).

\end{thebibliography}
\end{document}